\documentclass[%
 reprint,
 superscriptaddress,
nofootinbib,
 amsmath,
 amssymb,
 aps,
 pra,
]{revtex4-2}

\usepackage{graphicx}
\usepackage{dcolumn}
\usepackage{bm}
\usepackage{dsfont}
\usepackage{braket}
\usepackage{hyperref}
\usepackage{xcolor}
\usepackage{soul}
\usepackage{ulem}
\usepackage{float}
\usepackage{amsmath}
\usepackage{footnote}

\begin{document}

\preprint{APS/123-QED}

\title{Probing Majorana Modes via Local Spin Dynamics}

\author{Johannes Bjerlin}
\email{Johannes.Bjerlin@matfys.lth.se}
\affiliation{Niels Bohr Institute, University of Copenhagen }
\affiliation{Department of Physics and Astronomy, University of Southern California, Los Angeles, California 90089-0484, USA}
\affiliation{Mathematical Physics, Lund University }
\author{Anders S. S\o rensen}
\affiliation{Niels Bohr Institute, University of Copenhagen }
\author{Stephan Haas}
\affiliation{Department of Physics and Astronomy, University of Southern California, Los Angeles, California 90089-0484, USA}

\date{\today}

\begin{abstract}
We investigate Majorana modes in a quantum spin chain with bond-dependent exchange interactions by studying its dynamics.
Specifically, we consider two-time correlations for the Kitaev-Heisenberg (KH) Hamiltonian close to the so-called Kitaev critical point. 
Here, the model coincides with a phase boundary of two uncoupled instances of Kitaev's model for p-wave superconductors, together supporting a degenerate ground state characterized by multiple Majorana modes. In this regime, the real-time dynamics of local spins reveal a set of \textit{strong zero modes}, corresponding to a set of protruding
frequencies in the two-time correlation function.
We derive perturbative interactions that map the KH spin chain onto the topological regime of Kitaev's fermionic model, thus opening up a bulk gap whilst retaining almost degenerate modes in the mesoscopic regime, i.e., for finite system sizes. 
This showcases the emergence of Majorana modes in a chain of effective dimers.
Here, the binding energy within each unit cell competes with the inter-dimer coupling to generate a finite size energy gap, in analogy with local energy terms in the transverse-field Ising model. These modes give rise to long coherence times of local spins located at the system edges.
By breaking the local symmetry in each dimer, one can also observe a second class of Majorana modes in terms of a beating frequency in the two-time correlations function of the edge spin.
Furthermore, we develop a scenario for realizing these model predictions in ion-trap quantum simulators with collective addressing of the ions.
\end{abstract} 
\maketitle

\section{\label{sec-1Intro} Introduction}

Topological modes are ubiquitous in many-body (MB) models, but their experimental detection and control in naturally occurring quantum systems can be challenging~\cite{xu2015experimental}. A prominent example is the Majorana fermion (MF), a non-Abelian anyon with non-trivial exchange statistics~\cite{stern2010non}, which
has been studied for a wide range of MB systems \cite{leijnse2012introduction}. The perhaps simplest manifestation of an MF was proposed by Kitaev, who introduced a toy model for a fermionic quantum wire in the form of a one-dimensional (1D) p-wave paired superconductor~\cite{kitaev2001unpaired}. The microscopic origin of this model was worked out for $d^5$ transition metals~\cite{jackeli2009mott}, and it is up until this day an important tool in the active pursuit of controllable MFs \cite{agrapidis2018ordered}. 
Kitaev's fermionic model is intimately connected to the Ising~\cite{greiter20141d} and Kitaev-Heisenberg (KH) spin models. While the Ising model is ubiquitous and studied extensively in many contexts the KH model, with a potential realization in ${4d^5}$
ruthenium trichloride $\alpha$-RuCl3~\cite{kubota2015successive,banerjee2016proximate},  is less commonplace. The KH model attains frustration due to bond-dependent exchange couplings, and it may support long-range magnetic order~\cite{singh2010antiferromagnetic,ye2012direct} and quantum spin-liquid states (SQLs)~\cite{chaloupka2010kitaev,banerjee2016proximate}. Specifically in 1D the prospect for SQLs and topological modes has also been investigated~\cite{katsura2015exact,brzezicki2007quantum}.

For large-scale calculations of SQLs and topological modes, the use of quantum simulation with engineered lattice Hamiltonians in cold atom systems is a viable pathway \cite{schafer2020tools,cooper2019topological}. However, these systems typically require extremely low temperatures. A favorable alternative is given by simulators based on trapped ions~\cite{cirac1995quantum,leibfried2002trapped,wineland1998trapped}. Such setups are versatile and can function at comparatively high temperatures~\cite{sorensen1999quantum}. Recent successful examples of ion-trap simulations include a dynamical phase transition for a 53- qubits system~\cite{zhang2017observation}, as well as quasiparticle dynamics in an Ising spin chain~\cite{jurcevic2014quasiparticle}.
While quantum simulation promises remarkable speed up in the characterization of complex systems~\cite{boixo2018characterizing}, so far most implementations have focused on well-studied static properties for which other highly effective numerical and analytical tools are available.
Quantum simulation of dynamical features is hence especially compelling~\cite{xie,daug2020topologically}, as classical calculations are typically very costly~\cite{white2004real,anders2005real}. While there have been some advancements in classical computation of time-dependent observables~\cite{cohen2015taming},  there is to this date no general method to efficiently simulate the dynamics of large and strongly correlated systems. 

Recent examples of dynamical quantum simulation include studies of two-time correlations (TTCs)~\cite{gomez2016quantum,gomez2018universal,mendoza2019enhancing} and out-of-time correlations (OTOCs)~\cite{shen2017out,syzranov2018out} in interacting models, which have provided a new understanding of phase transitions and MB modes.
In addition, the Ising model was probed using the real-time dynamics of a single spin~\cite{gessner2014observing}. This type of dynamics has been studied for edge spins in a range of open boundary models. The spins here may exhibit long coherence times owing to the presence of strong zero modes~\cite{fendley2016strong,kemp2017long,jermyn2014stability}. Furthermore, fermionic models with topological Majoranas have been studied via survival rates of edge modes~\cite{wang2018detecting} and via Leggett-Garg inequalities~\cite{gomez2018universal}.
Such techniques are powerful since they can be used at high temperatures~\cite{kemp2017long,daug2020topologically,colbert2014proposal}. 

An interesting theme, adjacent to quantum simulation and condensed matter physics, concerns the territory of few-to-many body physics. Here, recent advances in computational and experimental techniques (particularly within ultracold atomic gases~\cite{serwane2011deterministic,bloch2008many}), has sparked experimental studies of, e.g., few-body magnetism without a lattice in one dimension~\cite{murmann2015antiferromagnetic}, the formation of a Fermi sea~\cite{wenz2013few} and, more recently, a few-body analogue of a quantum phase transition in two dimensions~\cite{bayha2020observing,bjerlin2016few}.

The advancements mentioned above highlight the growing interest in the controlled simulation of mesoscopic systems and number-conserving models with exotic features, which can shed light on the origins of quantum MB phases.
Already, several studies have been conducted on MFs and topological phases in number conserving lattice models,  motivated by the quest for a topological quantum computer~\cite{nayak2008non}.
This includes numerical studies of the topological features themselves, using density matrix renormalization group techniques~\cite{kraus2013majorana,iemini2015localized,agrapidis2018ordered}, as well as studies focusing on the microscopic origins and possible realizations of the models in which they arise~\cite{lang2015topological,zhang2017observation,kraus2013majorana,jiang2011majorana,sau2011number}.
This also extends to studies of dynamical observables~\cite{xie,jiang2011majorana,jermyn2014stability}, and
in a recent preprint some of the few-body aspects of Majorana quasiparticles were laid out~\cite{bland2020observability}, underlining the promise of quantum simulation of few-body physics as a way to study complex MB phenomena using a bottom-up approach.

Here we focus on few-body phases of an interacting 1D quantum spin model (SM) that emulates Majorana edge modes (MEMs), investigating its dynamical features in the few-to-many body limit. The term "emulate" refers to the fact that the MEMs are topologically non-trivial only in the fermionic representation of the model~\cite{greiter20141d}. Interestingly, the dynamical features of the SM still manifest a large discrepancy between bulk and edge. 
We begin by presenting an appropriate form of the Kitaev-Heisenberg Hamiltonian~\cite{agrapidis2018ordered}, using two parameters to tune the system between different phases around one of its critical points. We briefly discuss the various relevant phases in the static regime before investigating their individual dynamical signatures in local spin observables, focusing on MEMs. 
We use two-time correlation (TTC) functions to probe the Majorana bulk gap as well as the interaction-induced energy splitting 
between edge modes. This detection protocol elucidates the few-to-many-body development of MEMs without the requirement of deterministic preparation of any particular quantum state.
Finally, we discuss a possible experimental realization of these findings in an ion-trap setup.

The static properties of the Hamiltonian are studied by means of exact diagonalization, using the full basis set of $\hat{S}^\text{z}_\text{i}$ eigenstates. We utilize a sparse representation of the Hamiltonian and obtain the low-lying eigenvectors using the open-source library \textbf{Eigen}
~\cite{eigenweb} developed for \textbf{c++}. 
For determining dynamical features, we numerically solve the time-dependent Schrödinger equation, using the fourth-order Runga-Kutta method for temporal discretization. Here, the sparse matrix-vector multiplication can be easily parallelized and distributed over multiple cores. Using this setup, we can currently treat systems of up to chain lengths $L\sim 20$ on a single standard machine. 
\begin{figure*}
\includegraphics[width=0.7\textwidth]{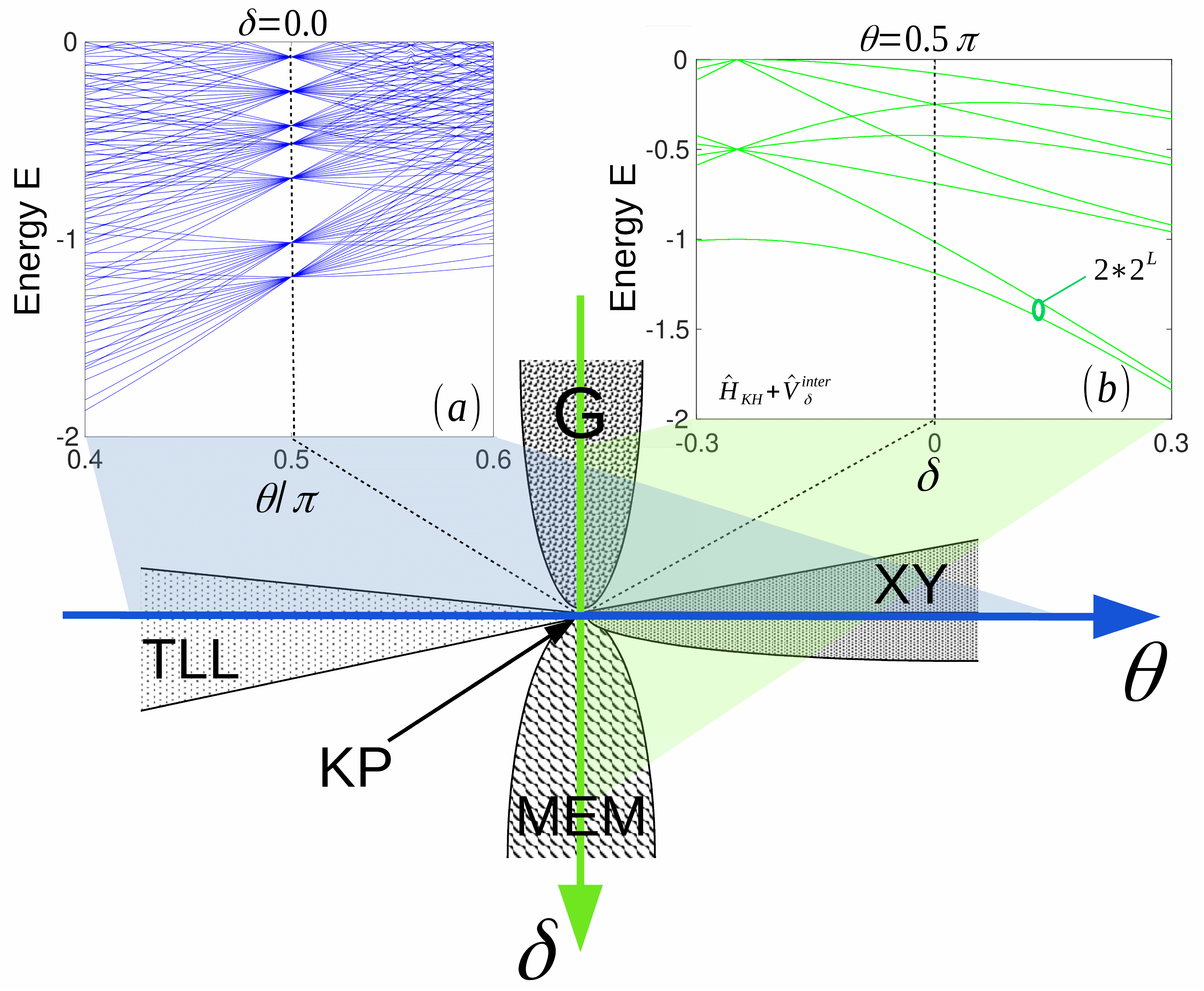}
 \caption{
 Phase diagram of the Kitaev-Heisenberg (KH) quantum spin model for $L=8$ in the vicinity of a critical Kitaev point (KP), located at ($\delta=0$, $\theta= \pi/2$). At this point, the Hamiltonian describes a fermionic p-wave superconductor that supports a degenerate ground state with multiple Majorana modes. The hierarchy of degenerate multiplets at the KP is directly observed in the spectrum in subplot $(a)$. Adding a term $ \hat{V}_\delta$ with strength $\delta> 0$ drives the system into the Majorana edge mode (MEM) phase, giving rise to a bulk gap, dividing all energy states into two sectors. Perturbing around the KP with the Y-bond interaction term $\hat{V}^{\text{inter}}_\delta$ in Eq.~\ref{pertB} realizes a dimer Majorana edge mode phase, with each level attaining a $2\cdot 2^\text{L}$ degeneracy for $L\rightarrow\infty$, whereas a non-vanishing  gap is attained for finite $L$, as seen in subplot (b).
 For $\theta=\pi/2- d\theta$ and $\delta\sim 0$, shown in subplot (a),  
 the system becomes a Tomonaga Luttinger liquid (TLL), which persists for moderate perturbation strengths $|\delta|$. For $\theta=\pi/2+d\theta$ and $\delta\sim 0$, the system is in a spiral XY phase, also persisting for moderate perturbations. For $\theta \sim \pi/2$ and $\delta <0$, the system is in a gapped (G) phase with no MEMs. 
 The MEM phases show a distinctly different behavior than the other phases in terms of the dynamical development of local spins.}\label{Fig1} 
\end{figure*}
%
%
%
%
%
%
%
%
%
\section{\label{sec-2Model} A Tunable Model for Majorana Edge Modes}
We focus on the 1D KH model describing an even number $L$ of spin-$1/2$ subsystems interacting via nearest neighbor (NN) couplings. The unit cells consist of two spins, where the interaction inside the unit cell is different from the interaction between neighboring unit cells. The interaction between the spins is described by the Hamiltonian
\begin{eqnarray}
\hat{H}_{KH}+\hat{V}_\delta  & = &
K\sum^\text{L/2}_{\text{j=1}} (\hat{S}^\text{x}_{\text{2j-1}} \hat{S}^\text{x}_{\text{2j}} +
    \hat{S}^\text{y}_{\text{2j}} \hat{S}^\text{y}_{\text{2j+1}}) \nonumber \\ \nonumber
    & + & J \sum^\text{L}_{\text{i=1}}  (\hat{S}^\text{x}_\text{i}\hat{S}^\text{x}_{\text{i+1}}  + \hat{S}^\text{y}_{\text{i}}\hat{S}^\text{y}_{\text{i+1}})+\hat{V}_{\delta} \\ 
& = &  \frac{2J+K}{4} \sum^\text{L}_{\text{i=1}} (\hat{S}^\text{+}_{\text{i}} \hat{S}^\text{-}_{\text{i+1}} +
    \hat{S}^\text{-}_{\text{i}} \hat{S}^\text{+}_{\text{i+1}}) \nonumber \\ 
& + & \frac{K}{4} \sum^\text{L}_{\text{i=1}}  
    (-1)^\text{i} (\hat{S}^\text{+}_\text{i}\hat{S}^\text{+}_{\text{i+1}}  + \hat{S}^\text{-}_{\text{i}}\hat{S}^\text{-}_{\text{i+1}})+ \hat{V}_{\delta}  
    \label{Hamiltonian}
\end{eqnarray}
Here $j$ is the unit cell index, and $i$ is the spin index. $\hat{S}^\text{x}_\text{i}$ corresponds to a local operator of spin $i$, describing spin along the $x$-axis. We further set $\hbar=1$, so that $\hat{S}^\text{D}_{i}=1/2\cdot\sigma^\text{D}_\text{i}$ in terms of Pauli matrices $\sigma_\text{i}^\text{D}$.
Initially neglecting the last term, $\hat{V}_{\delta}$, this quantum MB Hamiltonian, with tunable parameters $J$ and $K$, can be considered an inhomogeneous Heisenberg $XY$-model with exchange terms and additional sign-alternating double spin-flip interactions~\cite{kitaev2006anyons,agrapidis2018ordered}. Similar models have been studied in the context of quantum phase transitions, criticality and magnetic long-range order~\cite{mahdavifar2010numerical,eriksson2009multicriticality,brzezicki2007quantum}.
We parametrize it in terms of a polar parameter, $\theta$, governing the relative strength and signs of the interactions according to
\begin{equation}
\begin{split}
        & K=\sin{\theta},\\
        & J=\cos{\theta},
    \end{split}
\end{equation}
and we use $\sqrt{K^2 + J^2}=1$ as the unit of energy throughout.

We begin by studying the phase diagram of the system with $\hat{V}_{\delta}=0$ close  
to the so-called \textit{Kitaev points}, located at 
$\theta_\text{KP}= \pm \pi/2 \rightarrow K=\pm 1,J=0$ and $\theta_\text{KP}= 5\pi/4 \pm\pi/2 \rightarrow K=\pm 1/\sqrt{2},J=\mp 1/\sqrt{2}$. Applying the Jordan-Wigner transformation, we can find the corresponding fermionic model (see Supplemental materials~\ref{asec-1spectral}). The fermionic Hamiltonian can be directly decomposed into two separate systems, A and B, of length $L/2$, each corresponding to one instance of Kitaev's model for a $p$-wave paired superconductor~\cite{kitaev2001unpaired} at the boundary point between the trivial and topological phase (see Supplemental materials~\ref{asec-1spectral}). 
Exactly at the Kitaev points, only one of the subsystems A or B contributes energy in the Hamiltonian, so the full system acquires one free spin per unit cell, leading to groundstate degeneracies $2^\text{L/2}$ and $2^\text{L/2 -1}$ for open and closed chains, respectively~\cite{agrapidis2018ordered}. For the open chain this amounts to $L/2$ Majorana operators, which are entirely absent from the Hamiltonian, so that the entire spectrum exhibits the same degeneracies as found in the groundstate. This global degeneracy is a stronger condition than what is usually required for general topological order~\cite{alicea2016topological}.
The system here hosts multiple bulk Majorana modes distributed all across the chain, with a hierarchy of multiply degenerate states. Specifically, the highly degenerate groundstate multiplet is separated from the excited states by a gap, a necessary condition for the presence of non-Abelian quasiparticles~\cite{stern2010non,leijnse2012introduction}. 
Throughout the text, a \textit{globally} $N$-fold degenerate spectrum means that  each level in the spectrum is \textit{at least} $N$-fold degenerate, but additional degeneracies may be present.

In this work, we focus specifically on the realization of MEMs around the Kitaev point $\theta_\text{KP}=\pi/2$.
In Kitaev's original model, the MEM phase supports topologically protected modes at the edges~\cite{kitaev2001unpaired}, which correspond to a spontaneously broken spin-reflection symmetry when mapped to the Ising spin model~\cite{greiter20141d}. 
We will nevertheless use the term MEM also in the spin picture.

To achieve the MEM phase in our setup we must invoke the additional term $\hat{V}_{\delta}$ into the Hamiltonian.
Starting at the Kitaev point $\theta_\text{K}=\pi/2 \rightarrow K=1$ we map the system onto
\begin{equation}
    \hat{H}_\text{A}= \frac{K_1}{2}\sum_{\text{j=1}} \hat{d}^\dagger_\text{j}\hat{d}_\text{j} +\frac{K_2}{4}\sum_{\text{j=1}}(\hat{d}^\dagger_\text{j}\hat{d}_{\text{j+1}} + \hat{d}^\dagger_\text{j}\hat{d}^\dagger_{\text{1+j}}) + h.c.,
    \label{kitmodelA}
\end{equation}
with fermionic creation(annihilation) operators $\hat{d}^\dagger_\text{j}$($\hat{d}_\text{j}$). Here $K_1=K_2=1$, and $j$ is the unit cell index (see Supplemental materials~\ref{asec-1spectral}). 
Comparing to the Kitaev model~\cite{kitaev2001unpaired}, this gives the boundary point of the $p$-wave paired superconductor. Therefore, for topological modes the relative size of the first term must be decreased, so that $|{K_1}|<K_2$. We may thus either decrease $|{K_1}|$ or increase $K_2$ to enter the topological regime.

We first consider the (local) energy term proportional to $K_1$ and map this back to the spin picture (see Supplemental materials~\ref{asec_gappedmodederivation}), revealing the appropriate perturbation term, 
 \begin{equation} \hat{V}^{\text{intra}}_{\delta}= \delta \sum_{\text{j=1}}^\text{L/2} \hat{S}^\text{x}_{\text{2j-1}}\hat{S}^\text{x}_{\text{2j}},\label{pertA}
\end{equation}
with the MEM phase occuring for $|{K_1}+\delta|<K_2$. This term corresponds to interactions \textit{within} a unit cell of two spins.
We can compare this situation to the equivalence of the transverse-field Ising model and the Kitaev model~\cite{greiter20141d,backens2017emulating}, where the local fermionic energy term maps onto the local energy of a single spin in a magnetic field. 
For our case, each term in Eq.~\ref{pertA} instead represents the local energy of the unit cell dimer $j$.
Precisely at the Kitaev point, where $K_1=K_2=1$, the dimer energy equals that of the inter-dimer bond, and the system remains gapped for $L\rightarrow \infty$. Here, the global degeneracy is that of $L/2$ dimers with one free spin each, giving $2^\text{L/2}$ states. 
For $|{K_1}|<K_2$ the inter-dimer bonds instead dominate, and an additional global two-fold symmetry arises for $L\rightarrow\infty$, corresponding to zero-energy Majorana modes, giving a global degeneracy of $2 \cdot 2^\text{L/2}$. 
The degeneracy is perfect in the limit of infinite chains, whereas the finite size gap between the two degenerate multiplets scales with $e^\text{-L/2}$.
Aside from additional degeneracies, the energy spectrum of this system coincides perfectly with that of a transverse field Ising model with $L/2$ spins, $\hat{H}_{I}=\sum K_2\hat{S}^\text{y}_{\text{i}}\hat{S}^\text{y}_{\text{i+1}} + K_1\hat{S}^\text{x}_\text{i}$.
The Hamiltonian~\ref{Hamiltonian} is thus very similar to the transverse field Ising model but differs in its dynamical properties due to the additional degeneracies.

As noted above, we can also enter the MEM phase by increasing the relative size of the terms scaling with $K_2$ in the Hamiltonian~\ref{kitmodelA}, giving
 \begin{equation} \hat{V}^{\text{inter}}_{\delta}=\delta \sum_{\text{j=1}}^\text{L/2} \hat{S}^\text{y}_{\text{2j}}\hat{S}^\text{y}_{\text{2j+1}},\label{pertB}
\end{equation}
with the unit cell index $j$ and the MEM phase occurring for $|{K_1}|<K_2+\delta$.
This term corresponds to interactions \textit{between} two unit cells. 

We now invoke a third option for the perturbing interaction, $\hat{V}_\delta$, corresponding to a \textit{fully connected} Ising term,
 \begin{equation} \hat{V}^{\text{Ising}}_{\delta}=\delta\sum_{\text{i=1}}^\text{L} \hat{S}^\text{y}_{\text{i}}\hat{S}^\text{y}_{\text{i+1}},\label{pertC}
\end{equation} where $i$ is the spin site index. 
This perturbation does not map the fermionic Hamiltonian onto a Kitaev model, but we nevertheless see the emergence of an MEM phase for $|{K_1}|<K_2+\delta$.
We will see that this perturbation simultaneously creates MEMs and breaks local symmetries within each dimer, giving rise to a beating pattern in the time-dependent edge spin correlation functions.
In conclusion, we use $\delta$ as an effective parameter that controls the onset of MEMs, using either of the perturbations in Eq.~\ref{pertB} or Eq.~\ref{pertC}. The two different perturbations are used to highlight two different effects in dynamical simulations of the MEM regime. The spectrum due to the inter-dimer perturbation Eq.~\ref{pertB} is depicted in Fig.~\ref{Fig1}.

\section{Phase diagram and static properties}
Let us now briefly discuss the four phases in the phase diagram shown in Fig.~\ref{Fig1}, spanned by the parameters $\theta$ and $\delta$ in the vicinity of the critical Kitaev point at $\theta=\pi/2$ and $\delta=0$. The characterization of these phases will be helpful when discussing the dynamical features of local spins in the later sections.
\begin{itemize}
    \item Majorana Edge Mode (MEM) phase ($\theta = \theta_\text{KP}=\pi/2, \delta>0 $): 
    the bulk energy spectrum in this regime is gapped, with two zero-energy edge modes in the thermodynamic limit.
    However, in finite systems, their energies remain small but finite, yielding a finite-size gap that vanishes exponentially with increasing system size. For the perturbing term, $\hat{V}^{\text{inter}}_{\delta}$, each level has a global $2^\text{L/2}$-fold degeneracy due to free parameters in the Hamiltonian, so that for $L\rightarrow \infty $ the spectrum becomes $2 \cdot 2^\text{L/2}$-fold degenerate. We call this the \textit{dimer} MEM phase.
    For  $\hat{V}^{\text{Ising}}_{\delta}$ the spectrum becomes globally two-fold degenerate for $L\rightarrow \infty$. We call this the \textit{Ising} MEM phase.
    \item Gapped (G) phase ($\theta = \theta_\text{KP}=\pi/2, \delta \lesssim 0 $): this regime has a gapped energy spectrum, with no Majorana edge modes present.
    \item Spiral XY phase ($\theta >\theta_\text{KP}=\pi/2, \delta = 0$): the energy spectrum in this regime is gapless, and no Majorana modes are present.
    
    \item Tomonaga Luttinger Liquid (TLL) phase ($\theta < \theta_\text{KP}=\pi/2, \delta \approx 0 $): The low energy spectrum is gapless. 
\end{itemize}
 Further characterization and discussion of the static properties of these phases are presented in Supplemental materials~\ref{asec_phases}.
\section{Dynamical footprints of the Majorana modes}
Time-dependent observables are a powerful tool for the analysis of physical systems beyond their groundstate phases~\cite{kemp2017long,daug2020topologically, gomez2018universal,mendoza2019enhancing,gessner2014observing}.
In particular, local measurements, $\hat{S}^\text{D}_\text{i}(t)$, of a spin $i$ along $D\in \{ x,y,z \}$ are intuitive and experimentally accessible probes that can be used to highlight the emergence of MEMs~\cite{kemp2017long,else2017prethermal}.

We first consider the spin operator,
\begin{equation}
    \hat{\mathcal{G}}^\text{D}= \prod^{\text{L}}_{\text{k=1}}\hat{\sigma}^\text{D}_{\text{k}}.
    \label{spinflip}
\end{equation}
The eigenstates of $\hat{\mathcal{G}}^\text{D}$ are denoted $|\pm \Phi^\text{D}_\text{n} \rangle=\left[| {s^\text{D}_1} \smallskip s^\text{D}_2 \smallskip s^\text{D}_3 ... \rangle \right]$, where $\pm$ denotes positive or negative parity, respectively. These eigenstates will serve as the initial states for the dynamical simulations, where we numerically evolve each state in time under the Hamiltonian operator~\ref{Hamiltonian} and study the dynamical evolution of local spins.
We also note that this spin-operator flips all spins along the axes perpendicular to $D$, i.e., along $x$ and $y$ for $\hat{\mathcal{G}}_\text{z}$.

\begin{figure*}
\includegraphics[width=0.8\textwidth]{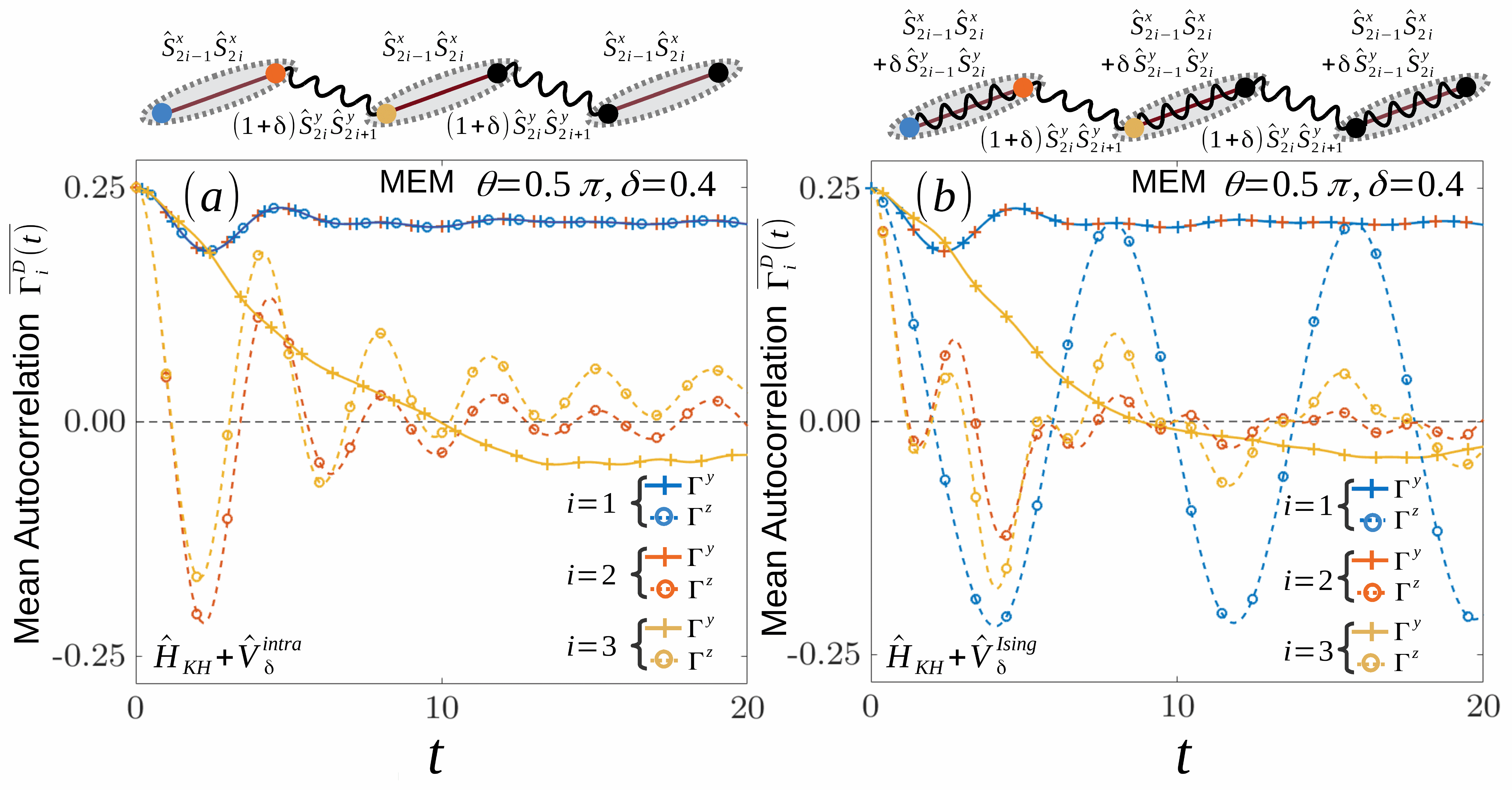}
 \caption{Time evolution of mean autocorrelation $\overline{\Gamma^\text{D}_\text{i}(t)}$ at an edge site, $i=1$, and at a bulk sites, $i=2$,$i=3$ for a system of $L=14$ spins. 
 (a) shows the mean autocorrelation function in the MEM phase achieved for $\hat{H}^{\text{inter}}_\delta=\hat{H}_\text{KH}(\theta=\pi/2)+\hat{V}^{\text{inter}}_\delta$ with $\delta=0.4$. 
Both $i=1$ and $i=2$ show long-time coherence for a spin initially pointing in the $y$-direction, owing to the presence of zero modes which commutes with $\hat{H}$ up to an exponentially small factor $\epsilon$. This is also true for spin-$z$, but only at the edge.
Other spins decay rapidly.
(b) shows the corresponding autocorrelations for $\hat{H}^{\text{Ising}}_\delta=\hat{H}_\text{KH}(\theta=\pi/2)+ \hat{V}^{\text{Ising}}_\delta$. Here the autocorrelation $\overline{\Gamma^\text{z}_\text{i}(t)}$ for an edge spin $i=1$ along $z$ oscillates perpetually with an amplitude scaling with the corresponding value of $\overline{\Gamma^\text{y}_{1}(t)}$.    }
 \label{Fig2a} 
\end{figure*}

For quantitative measures, we consider the mean autocorrelation function,
\begin{equation}
\overline{\Gamma^\text{D}_\text{i}(t)}=(1/N) \sum^\text{N}_\text{n} \Gamma^\text{D}_\text{i}(t)=(1/N)\sum^\text{N}_\text{n} \langle \hat{S}^\text{D}_\text{i}(t) \hat{S}^\text{D}_\text{i}(t=0)\rangle_\text{n},
\label{autocorrelation}
\end{equation}
where the sum over N produces the average over a randomly sampled set of $N$ initial states $| \pm\Phi^\text{D}_\text{n} \rangle \in \{ | \downarrow \downarrow \downarrow ... \rangle,..., |\downarrow \downarrow \uparrow ...\rangle$\} in the basis of spin $D$. 
We also consider the (discrete) Fourier transformed evolution functions  $\mathcal{F}\left( \langle \hat{S}^\text{D}_\text{i} (t) \rangle \right)$, again taking the average over a large set of initial states,
\begin{equation}
    |c^\text{D}_\text{i}(\omega)|= (1/N) \left| \sum^\text{N}_\text{n}  \mathcal{F}\left( \langle \hat{S}^\text{D}_\text{i} (t)\rangle_\text{n} \right)\right|.
    \label{frequency}
\end{equation}
We further calculate variances to highlight which features are largely independent of the particular input states we choose.\footnote{Because of computational limitations, the 
frequency-dependent quantities
are generally displayed on rather coarse grids in frequency. We stress however that the significant features, which will be used to identify the MEM phases, are visible already for rather short evolution times. 
To extract more detailed information we run simulations for longer times.}
By sampling over multiple initial states and taking the average, we specifically access robust features of the system in the sense that an experimental setup would not 
rely on repeated and deterministic preparation of any specific initial state. Measurements can instead be performed with mixed states for those spins which are not directly probed, which is especially relevant for detection of \textit{strong zero modes}~\cite{kemp2017long,jermyn2014stability}. To simplify the computations we, however, consider pure initial states for the individual runs and take the average afterward, i.e. we essentially perform a Monte-Carlo sampling of a completely mixed density matrix. 
Fig.~\ref{Fig2a} shows mean autocorrelations for a set of randomly sampled states developing in time under two different Hamiltonians. As will be discussed in the following section, there are several robust features in the mean autocorrelations of edge spins (like the constant spin-$y$ projection in each plot), even though they essentially represent time-development of mixed states.

\subsection{Zero modes of the Hamiltonian \label{Zeromodes}}
A Hamiltonian that supports MEMs can be represented in terms of Majorana operators in such a way that some of them drop out of the Hamiltonian in the infinite ($L\rightarrow \infty$) system limit~\cite{kitaev2001unpaired}, giving rise to global symmetries and corresponding degeneracies in the entire energy spectrum. 
In line with the procedures in Refs.~\cite{kitaev2001unpaired,fendley2016strong,kemp2017long}, we elucidate the dynamical properties of the finite-size system by first deriving the zero modes, which correspond to the Majorana modes in the corresponding fermionic model. These modes are constructed to approximately commute with the Hamiltonian, with corrections $\sim e^{-\text{L}}$, and are associated with long-time coherent features in the dynamical evolution. 

We first consider the Hamiltonian~\ref{Hamiltonian} at the Kitaev point $\theta=\pi/2$, using the perturbation $\hat{V}^{\text{inter}}_\delta$, so that
\begin{equation}
\begin{split}
 \hat{H}^{\text{inter}}_{\delta}=& \hat{H}_{\text{KH}}(\theta=\pi/2) +\hat{V}^{\text{inter}}_{\delta} \\
 =&\sum^\text{L/2}_{\text{j=1}} (\hat{S}^\text{x}_{\text{2j-1}} \hat{S}^\text{x}_{\text{2j}} +  \hat{S}^\text{y}_{\text{2j}} \hat{S}^\text{y}_{\text{2j+1}}) +\delta \sum_{\text{j=1}}^\text{L/2} \hat{S}^\text{y}_{\text{2j}}S^\text{y}_{\text{2j+1}}\\ \equiv & (1+\delta)\left(\hat{H}_0 + \hat{W}\right).
 \label{HDMEM}
   \end{split}
 \end{equation}
For convenience, we have rescaled the Hamiltonian in the last line, so we end up with
\begin{equation*}
\begin{split}
&\hat{H}^\text{inter}_\delta \rightarrow \hat{H}_0 + \hat{W},\\
& \hat{H}_0= \sum^\text{L/2}_{\text{j=1}}\hat{S}^\text{y}_{\text{2j}} \hat{S}^\text{y}_{\text{2j+1}}, \\
& \hat{W}=\frac{1}{1+\delta}\sum^\text{L/2}_{\text{j=1}} \hat{S}^\text{x}_{\text{2j-1}} \hat{S}^\text{x}_{\text{2j}},
\label{hamiltonian_scaled}
\end{split}
\end{equation*}
where we recognize that $\delta>0$ yields the dimer MEM phase. 

We initially aim to construct a zero mode corresponding to a local spin along $y$ at the edge of the spin chain. 
This means that we simply put for the zeroth-order approximation of the mode operator $\Psi^{(0)}_\text{A}=\hat{S}^\text{y}_1$, which commutes with the
dominant term $\hat{H}_0$ in the Hamiltonian~\ref{HDMEM}.
It does, however, not commute with the full Hamiltonian, and for $\hat{H}_0+\hat{W}$ we find
\begin{equation*}
\left[ \hat{H}^{\text{inter}}_{\delta}, \Psi^{(0)}_\text{A} \right]=i\frac{1}{1+\delta}\hat{S}^\text{z}_{\text{1}}\hat{S}^\text{x}_{\text{2}},
\end{equation*}
which we offset by introducing first and second order terms $\Psi^{(1)}_\text{A}=\mathcal{M}_1\hat{S}^\text{z}_1\hat{S}^\text{y}_2$ and $\Psi^{(2)}_\text{A}=-4(1/(1+\delta))\hat{S}^\text{z}_1\hat{S}^\text{z}_2\hat{S}^\text{y}_3$, so that
\begin{equation*}
\begin{split}
\left[\hat{H}_0, \Psi^{(1)}_\text{A}+\Psi^{(2)}_\text{A} \right]&=-4\frac{1}{1+\delta}\left[\hat{S}^\text{y}_{\text{2}} \hat{S}^\text{y}_{\text{3}} , \Psi^{(2)}_\text{A} \right] \\ &=-i\frac{1}{1+\delta}\hat{S}^\text{z}_{\text{1}}\hat{S}^\text{x}_{\text{2}},
\end{split}
\end{equation*}
where the factor of four is absorbed by an emerging operator $\hat{S}^\text{y}_{\text{3}}\hat{S}^\text{y}_{\text{3}}=1/4$.
The first-order term actually commutes with the full Hamiltonian, so we can freely choose the constant $\mathcal{M}_1$.
$\Psi^{(2)}_\text{A}$ does not commute with $\hat{W}$, which again can be offset by additional terms of higher order.
Continuing in this fashion, we get
\begin{equation}
\begin{split}
\Psi_\text{A}&=
\sum^\text{L/2}_{\text{j=1}} \Psi_\text{A}^\text{(2j)}+\sum^\text{L/2}_{\text{j=2}}\Psi_\text{A}^\text{(2j-1)}\\
&=\mathcal{N}_{\text{e}} \sigma_1^\text{y} +\mathcal{N}_{\text{e}}\sum^\text{L/2}_{\text{j=2}} \left(-\frac{1}{1+\delta}\right)^\text{j-1} \sigma^\text{y}_{\text{2j-1}} \prod^{\text{2j-2}}_{\text{k=1}}\sigma^\text{z}_{\text{k}}\\ 
&+\mathcal{N}_{\text{o}}\sum^\text{L/2}_{\text{j=1}}  \mathcal{M}^\text{j-1}\sigma^\text{y}_{\text{2j}} \prod^{\text{2j-1}}_{\text{k=1}}\sigma^\text{z}_{\text{k}},
\label{Zmode}
\end{split}
\end{equation}
which we can show commutes with the Hamiltonian up to an exponentially small factor.
In this expression, we have changed the representation to Pauli spin operators, $\hat{S}^\text{D}_\text{i} \rightarrow \sigma^\text{D}_\text{i}/2$, and separated the odd and even orders for convenience. Each term in the sums now corresponds to a Majorana fermion~\cite{kitaev2001unpaired}.
For the second sum we have one free choice for the constant $\mathcal{M}_i$ per term, meaning each unit cell adds a degree of freedom for the zero mode. This is consistent with the global $2^\text{L/2}$-fold degeneracy in the spectrum. For simplicity, we have chosen $\mathcal{M}_j= \mathcal{M}^\text{j-1}$.

Each term in the sum of Eq.~\ref{Zmode} anticommutes with all the others, so that
\begin{equation}
\begin{split}
\Psi^2_\text{A}=& \mathcal{N}_{\text{e}}^2  \frac{  1- \left(\frac{1}{1+\delta} \right)^\text{L} }{ 1- \left( \frac{1}{1+\delta} \right)^2  } +\mathcal{N}_{\text{o}}^2  \frac{  1- \mathcal{M}^\text{L} }{ 1- \mathcal{M}^2   }\\ \approx& \mathcal{N}_{\text{e}}^2  \frac{  1 }{ 1- \left( \frac{1}{1+\delta}\right)^2   } +\mathcal{N}_{\text{o}}^2 \frac{  1 }{ 1- \mathcal{M}^2   } , 
\end{split}
\end{equation}
and we proceed to choose the normalization so that $\Psi^2_\text{A}=1$ for $|\mathcal{M}|<1$ and $\delta>0$.
It is now clear the each of the zero modes constructed in Eq.~\ref{Zmode} commutes with the Hamiltonian $H$, now in matrix representation, up to an exponentially small term,
\begin{equation}
 \begin{split}
\left[ \hat{H}^{\text{inter}}_{\delta}, \Psi_\text{A} \right]=& \left[ \frac{\sigma_{\text{L-1}}^\text{x} \sigma_\text{L}^\text{x}}{1+\delta}  , \mathcal{N}_{\text{e}} \left(-\frac{1}{1+\delta}\right)^{{\frac{L}{2}-1}} \sigma^\text{y}_{\text{L-1}} \prod^{\text{L-2}}_{\text{k=1}}\sigma^\text{z}_{\text{k}} \right]\\=&2i\frac{\mathcal{N}_{\text{e}}}{1+\delta}  \left(-\frac{1}{1+\delta}\right)^{\frac{L}{2}-1}
\left( \prod^{\text{L-1}}_{\text{k=1}}\sigma^\text{z}_{\text{k}}\right)\sigma^\text{x}_\text{L}\\
=&-2 \frac{\mathcal{N}_{\text{e}}}{1+\delta} \left(-\frac{1}{1+\delta}\right)^{\frac{L}{2}-1}\mathcal{G}^\text{z}
\sigma^\text{y}_\text{L}=\varepsilon_\text{rem},
\end{split}
\end{equation}
where $\mathcal{G}^\text{z}$ is the spin-flip operator in Eq.~\ref{spinflip}.
We also note that the zero mode anticommutes with the spin-flip operator so that $\{ {\mathcal{G}^\text{z}}, \Psi_\text{A} \} =0$.
This means that $\mathcal{G}^\text{z}$ toggles between different eigenstates of $\Psi_\text{A}$ and vice versa.
Together with normalizability and the vanishing commutator $\varepsilon_\text{rem}$, these properties of the zero mode constitute the necessary conditions for long-time edge spin coherence~\cite{kemp2017long}.
Alternatively, zero modes can be constructed similarly by starting at the other edge, $\hat{S}^\text{y}_\text{L}$.

We also find that for $\hat{S}^\text{y}_2$ and $\hat{S}^\text{z}_1$, which also commute with $\hat{H}_0$, two corresponding zero modes, $\Psi_\text{B}$ and $\Psi_\text{C}$ can be found (see Supplemental materials~\ref{asec_zeromodes}). These modes differ in their construction, since they are each derived from different starting points.
However, the different zero modes $\Psi_\text{A}$, $\Psi_\text{B}$, and $\Psi_\text{C}$ each span the same operator space, so that the implicated global degeneracy of the energy spectrum  for $L\rightarrow\infty$ is still $2 \cdot 2^\text{L/2}$. 
In summary, we can construct normalizable zero modes $\Psi_\text{A}$,$\Psi_\text{B}$, and $\Psi_\text{C}$ for the Hamiltonian~\ref{HDMEM}, which includes the perturbation  $\hat{V}^{\text{inter}}_\delta$ with $\delta>0$.

Interestingly, the zero modes $\Psi_\text{A}$ and $\Psi_\text{B}$ also commute, up to an exponentially small factor, with the Hamiltonian
\begin{equation}
\hat{H}^{\text{Ising}}_{\delta}= \hat{H}_{\text{KH}}(\theta=\pi/2) +\hat{V}^{\text{Ising}}_{\delta},
\label{Ising}
\end{equation}
which instead uses the perturbing Ising term $\hat{V}^{\text{Ising}}_{\delta}$.  This is true for $\delta>0$, providing we set all constants in front of odd orders to zero, so that $\mathcal{M}_\text{A}=0$ and $\mathcal{M}_\text{B}=0$. This means that for each of the operators, $\hat{S}_1^\text{y}$,$\hat{S}_2^\text{y}$, one may use the same construction of zero modes for Hamiltonian $\hat{H}^{\text{Ising}}_\delta$ as for $\hat{H}^{\text{inter}}_{\delta}$. 
For $\hat{H}^{\text{Ising}}_{\delta}$ the free parameters in the zero modes are however removed, so the global degeneracy in the spectrum becomes only two-fold in the limit $L\rightarrow\infty$.
The commutations between the finite-size zero modes and the Hamiltonians $\hat{H}^{\text{Ising}}_{\delta}$ and $\hat{H}^{\text{inter}}_{\delta}$ are identical, so the finite-size gap between zero modes $\Delta_\text{L}$ are also the same for the different perturbations.
We will see that the long-time coherence depends crucially on this gap, and we therefore expect some identical long-time features for both Hamiltonians.   
\subsection{Long time dynamics of edge spins \label{zeromodedynamics}}
We now proceed to summarize the impact of zero modes on the long-time dynamics of our system. For details, we refer the reader to the Supplemental materials~\ref{asec_zeromodedynamics}, and for a comprehensive theoretical background to Ref.~\cite{kemp2017long,fendley2016strong}.
We evaluate the autocorrelation function $ \Gamma^\text{D}_{1}(t)$ for an eigenstate $|S^\text{D}\rangle$, with corresponding eigenvalue $s^\text{D}_1$, of the edge spin operator $\hat{S}^\text{D}_1$ along direction $D$. In practice we will use states $|\pm \Phi^\text{D}_\text{n} \rangle$ which are eigenstates of $\mathcal{G}^\text{D}$, but here we consider a general state $|S^\text{D} \rangle$. We get for the autocorrelation 
\begin{equation}
\begin{split}
    \Gamma^\text{D}_{1}(t)=& \langle S^\text{D} |\hat{S}^\text{D}_{1}(t) \hat{S}^\text{D}_{1}(t=0)| S^\text{D} \rangle\\=& \frac{s^\text{D}_1}{2}\sum_\text{n,m} e^{-it\left( E_\text{m}-E_\text{n}\right)} \langle S^\text{D}|n\rangle \langle n| \sigma^\text{D}_1|m\rangle \langle m| S^\text{D} \rangle
\label{autocorr_bare}
\end{split}
\end{equation}
where $ \langle n|,\langle m|$ are eigenstates of some Hamiltonian $\hat{H}$ with corresponding zero mode $\Psi_\text{A}$.
Since $\Psi_\text{A}$ (almost) commutes with the Hamiltonian we may divide all energy states into two sectors denoted by positive or negative sign, corresponding positive or negative eigenvalues of $\Psi_\text{A}$ so that
\begin{equation}
\begin{split}
    &\hat{H}|n^\text{A}_{\pm}\rangle \approx E_\text{n}|n^\text{A}_{\pm}\rangle\\
    &\Psi_\text{A}|n^\text{A}_{\pm}\rangle= \pm|n^\text{A}_{\pm} \rangle
    \label{psi_commutation}
\end{split}
\end{equation}
We can now re-write the autocorrelation function with new indicies
\begin{align*}
    \Gamma^\text{D}_{1}(t)&= s^\text{D}_1 \sum_{\text{n}^\text{A},\text{m}^\text{A}} e^{-i\left(E_{\text{m}^\text{A}}-E_{\text{n}^\text{A}}\right)t} \\
    \cdot&\left( \langle S^\text{D} \left(|n^\text{A}_\text{+}\rangle \langle n^\text{A}_\text{+}|+|n^\text{A}_\text{-}\rangle \langle n^\text{A}_\text{-}|\right)\right.
   \\ \cdot& \left. \sigma^\text{D}_1\left(|m^\text{A}_\text{+}\rangle \langle m^\text{A}_\text{+}|+|m^\text{A}_\text{-}\rangle \langle m^\text{A}_\text{-}|\right) S^\text{D} \rangle \right)\\
\end{align*}
For long times $t$ and large system size $L$, terms with $n^\text{A}\neq m^\text{A}$ add up incoherently while terms with $n^\text{A}=m^\text{A}$ add up coherently, i.e. terms with $n^\text{A}\neq m^\text{A}$ get a random phase so that we can ignore them.  
The double sum may then be approximated for long times by
\begin{equation}
\begin{split}
\Gamma^\text{D}_{1}(t) \approx s^\text{D}_1 \sum_{\text{n}^\text{A}} \cdot \left(T^\text{D}_\text{1}+T^\text{D}_\text{2}+T^\text{D}_\text{3}+T^\text{D}_\text{4}\right)
\end{split}
\label{autocorr}
\end{equation}
with terms $T^\text{D}_1$-$T^\text{D}_4$ relating to the time-independent matrix elements between  $|n^\text{A}_\pm \rangle$.
\begin{equation}
\begin{split}
    T^\text{D}_1=\langle S^D |n_{A-} \rangle \langle n_{A-}| \sigma^D_1 |n_{A-} \rangle \langle n_{A-}|S^D\rangle  \\
        T^\text{D}_2=\langle S^D |n_{A-} \rangle \langle n_{A-}| \sigma^D_1 |n_{A+} \rangle \langle n_{A+}|S^D\rangle  \\
            T^\text{D}_3=\langle S^D |n_{A+} \rangle \langle n_{A+}| \sigma^D_1 |n_{A-} \rangle \langle n_{A-}|S^D\rangle  \\
                T^\text{D}_4=\langle S^D |n_{A+} \rangle \langle n_{A+}| \sigma^D_1 |n_{A+} \rangle \langle n_{A+}|S^D\rangle  \\
\end{split}
\label{terms_main}
\end{equation}
This shows that for an infinite system, the long-time spin oscillations are stable. For finite systems this is no longer the case, and the oscillations will eventually decay.
The coherence time, i.e. the time during which the spin autocorrelation function remains stable, either displaying a finite value or a persistent oscillation, is generally set by the
commutation between the Hamiltonian and the zero mode, which vanishes with $L\rightarrow\infty$~\cite{kemp2017long}. 
Interestingly, if the finite size gaps $\Delta_\text{L}$ between semi-degenerate states in a systems' spectrum are \textit{all} identical, spin autocorrelations which first appear to decay will have a revival time of order $1/\Delta_\text{L} \propto 1/\langle \varepsilon_\text{rem} \rangle$. 
If one other hand the gaps are different we will only have partial revivals. 
For integrable systems, like the Ising model, these revival times may be directly calculated~\cite{fendley2014free}. This is also true for the Hamiltonian in Eq.~\ref{HDMEM}, which will be apparent from the dynamical simulations. 

\subsubsection{Coherence for $\sigma^\text{y}_1$,$\sigma^\text{y}_2$ and $\sigma^\text{z}_1$}
So far we have not specified the direction $D$ in which we aim to measure the spin, and we proceed to study the effect of two particular choices of $\hat{S}^\text{D}_1$ based on the time-independent terms in Eq.~\ref{autocorr}.
Since we specifically use the Pauli-spin representation of operators in this chapter, we will use $\sigma^\text{D}_\text{i}$ to represent a spin operator at site $i$ along $D$.
We begin with the Hamiltonian $\hat{H}^{\text{inter}}_\delta$ and $\sigma^\text{y}_1$ with corresponding zero mode $\Psi_\text{A}$. For $T^\text{y}_2$ and $T^\text{y}_3$ we exploit the spin-flip operator in Eq.~\ref{spinflip} which anticommutes with $\Psi_\text{A}$ so that $\mathcal{G}^\text{z} |n^\text{A}_{\pm} \rangle =  |n^\text{A}_{\mp} \rangle$ and we get
\begin{equation}
\begin{split}
    T^\text{y}_2=&\langle S^\text{y} |n^\text{A}_\text{-} \rangle \langle n^\text{A}_\text{+}|S^\text{y}\rangle \cdot  \langle n^\text{A}_\text{+}|\{ \sigma_1^\text{y},\mathcal{G}^\text{z}\}|n^\text{A}_\text{+} \rangle /2\\
    T^\text{y}_3=&\langle S^\text{y}|n^\text{A}_\text{+} \rangle \langle n^\text{A}_\text{-}|S^\text{y}\rangle \cdot  \langle n^\text{A}_\text{+}| \{ \sigma_1^\text{y}, \mathcal{G}^\text{z}  \} |n^\text{A }_\text{+} \rangle/2\\
\end{split}
\label{T2T3}
\end{equation}
Now we note the anticommutation $\{\sigma^\text{y}_1,\mathcal{G}^\text{z} \}=0$, leading to $T^\text{y}_2=T^\text{y}_3=0$.
For $T^\text{y}_1$ and  $T^\text{y}_4$ we instead employ Eq.~\ref{psi_commutation}
\begin{equation}
\begin{split}
&T^\text{y}_1=|\langle S^\text{y} |n^\text{A}_\text{-} \rangle |^2 \cdot  
   \langle n^\text{A}_\text{+}|\{ \Psi_\text{A}, \mathcal{G}^\text{z} \sigma^\text{y}_1 \mathcal{G}^\text{z} \} |n^\text{A}_\text{+} \rangle/2 \\
&T^\text{y}_4=  |\langle S^\text{y} |n^\text{A}_\text{+} \rangle |^2 \cdot \langle n^\text{A}_\text{+}| \{ \Psi_\text{A}, \sigma^\text{y}_1  \} |n^\text{A}_\text{+} \rangle/2
\end{split}
\label{T1T4}
\end{equation}
 For $T^\text{y}_4$ we may use Eq.~\ref{Zmode} directly, giving $\{\Psi_\text{A},\sigma^\text{y}_1 \}=\mathcal{N}_\text{e}+\mathcal{C}$, where $\mathcal{C}$ represents (exponentially) small corrections. For $T^\text{y}_1$ we see that $\{\Psi_\text{A},\mathcal{G}^\text{z}\sigma^\text{y}_1\mathcal{G}^\text{z} \}=\{\Psi_\text{A},\sigma^\text{z}_1\sigma^\text{y}_1\sigma^\text{z}_1 \}=-\{\Psi_\text{A},\sigma^\text{y}_1\}$.
This gives  the long-time limit of the autocorrelation
\begin{equation}
    \Gamma^\text{y}_{1}(t) \approx s^\text{y}_1 \left(\mathcal{N}_\text{e}+ \mathcal{C}\right) \sum_{\text{n}^\text{A}}\left(|\langle S^\text{y} |n^\text{A}_\text{+} \rangle |^2  -|\langle S^\text{y} |n^\text{A}_\text{-} \rangle |^2\right)
\end{equation}
which only depends on the initial state and how much overlap it has with each sector of eigenstates for $\Psi_\text{A}$.
The exact form of the corrections $\mathcal{C}$ depend specifically on the model~\cite{kemp2017long}, but they are always exponentially decreasing with $L/2$.

We can use an identical derivation for $\sigma^\text{y}_2$ by making the substitutions $\Psi_\text{A}\rightarrow\Psi_\text{B}$ and $|n^\text{A}_{\pm} \rangle \rightarrow |n^\text{B}_{\pm}\rangle$. For $\sigma^\text{z}_1$ we instead put $\mathcal{G}^\text{z}\rightarrow \mathcal{G}^\text{x}$, $\Psi_\text{A}\rightarrow\Psi_\text{C}$ and $|n^\text{A}_{\pm} \rangle \rightarrow |n^\text{C}_{\pm}\rangle$.

Fig.~\ref{Fig2a} a shows the simulated dynamical development of spins for  $\hat{H}^{\text{inter}}_\delta$, confirming that the mean autocorrelation for $\sigma^\text{y}_1$,$\sigma^\text{y}_2$ and $\sigma^\text{z}_1$ is long-lived compared to other spins. This is directly explained by the fact that we can construct corresponding zero modes, as shown in the previous section. This is not true for the other operators shown in the plot, where the autocorrelation vanishes for long times.

\subsection{Beating patterns for edge spins\label{sec_beating}}
In Fig.~\ref{Fig2a} we study how the autocorrelation function compares for the Hamiltonians $\hat{H}^{\text{inter}}_\delta$ and $\hat{H}^{\text{Ising}}_\delta$.
The results for $\sigma^\text{y}_1$ and $\sigma^\text{y}_2$ are the same for the different Hamiltonians, whereas $\sigma^\text{z}_1$ is strikingly different. Curiously,  long time coherence is still present for $\hat{H}^{\text{Ising}}_\delta$, but with an oscillating factor which we find is independent of system size. The coherence time of the oscillation is however set by system size, as for $\sigma^\text{y}_1$ and $\sigma^\text{y}_2$. 

We can relate this result directly to the zero modes. We derive in the Supplemental materials~\ref{asec_spinbeating} that 
\begin{equation}
\begin{split}
     \Gamma^\text{z}_{1}(t) &\approx \frac{s^\text{z}_1}{4}  (\cos^2{\delta t} - \sin^2{\delta t})\left(\mathcal{N}_\text{e} + C\right) \\& \cdot \sum_{\text{n}^\text{C}} \left(|\langle S^\text{z} |n^\text{C}_\text{+}\rangle |^2-|\langle S^\text{z} |n^\text{C}_\text{-} \rangle |^2\right)
     \label{beating}
\end{split}
\end{equation}
We see that the expression by symmetry is, except for the time-dependent factor, identical to the autocorrelation $\Gamma^\text{y}_{1}(t)$, but here for an initial state $| S^\text{z} \rangle$. This precession of the edge spin $\sigma_1^\text{z}$ is hence given by an oscillation, with frequency $\delta$, and an envelope function given by the coherence time of $\sigma_1^\text{y}$.

%
%

\begin{figure*}
 \includegraphics[width=0.8\textwidth]{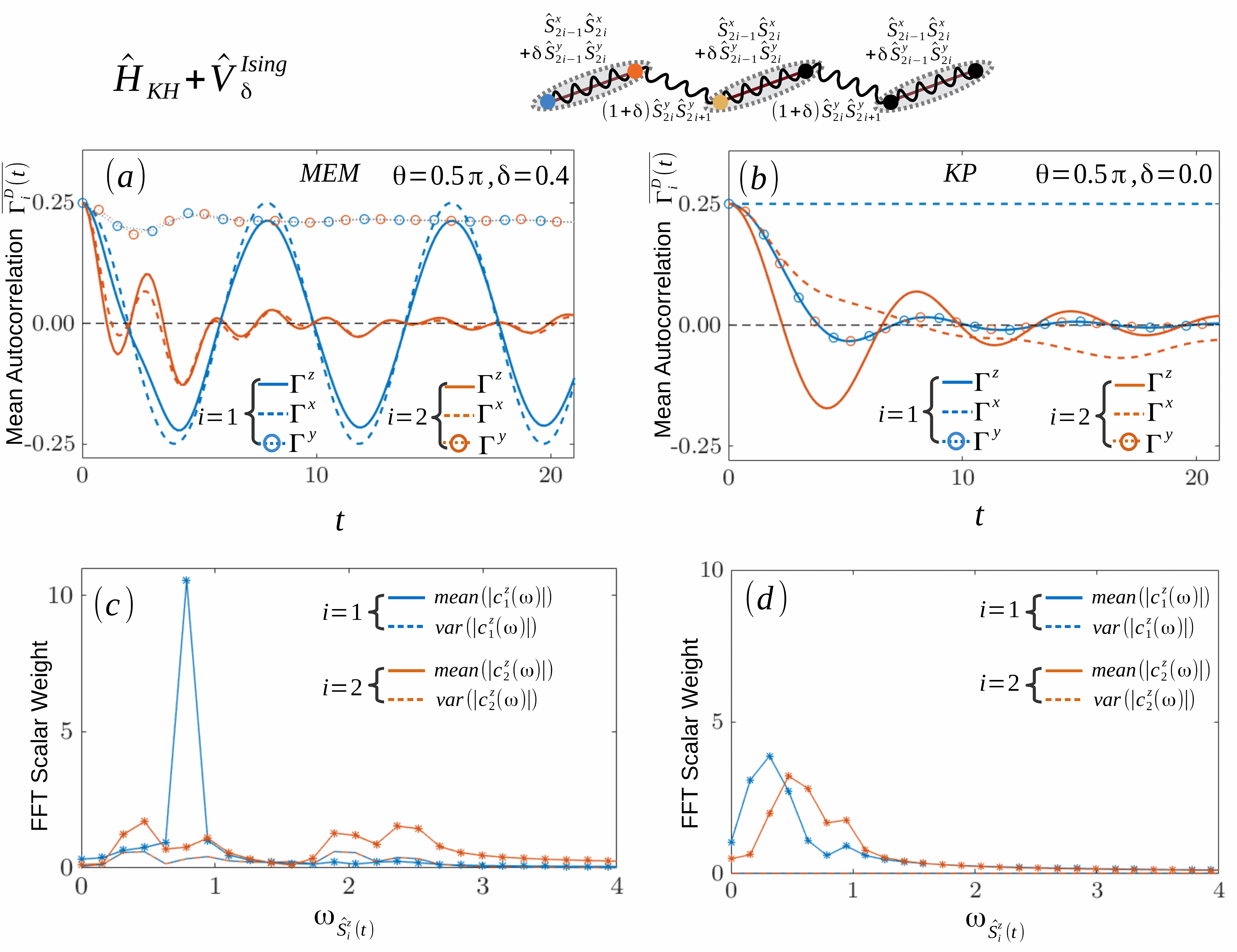}
 \caption{Time evolution of mean autocorrelation $\overline{\Gamma^\text{D}_\text{i}(t)}$ at an edge site, $i=1$, and at a bulk site, $i=2$ for a system of $L=14$ spins.  
 (a) and (b) show the mean autocorrelation function Eq.~\ref{autocorrelation} in the Ising MEM phase and at the Kitaev point, respectively. The former has Majorana modes only at the edges, and the latter has one Majorana mode per unit cell throughout the bulk. In the MEM phase, a single dominant oscillation frequency is observed for the edge site. This spin precession is only weakly damped, whereas oscillations in the bulk decay fast.
 At the critical point, (b), no characteristic frequency is observed.
 (c) and (d) show the frequency Fourier components Eq.~\ref{frequency} of the local spin-expectation value, $\langle \hat{S}^\text{z}_\text{i} (t) \rangle$, averaged over the same set of initial states as for the autocorrelation functions, for the MEM phase (c) and the Kitaev point (d).
 Shown are also the corresponding variances.
 The dominant single peak for the edge site $i=1$ has low relative variance and indicates a single oscillation frequency, which is independent of system size, $L$.
 The bulk site $i=2$ has a strikingly different behavior, with several oscillation frequencies and higher relative variances. The details of the frequency profile are dependent on system size, indicative of a bulk mode. 
 (d) shows the characteristic behavior of the Kitaev point, with a multi-peaked structure and zero variance at both edge and bulk sites.}
 \label{Fig2b} 
\end{figure*}
\begin{figure*}
\includegraphics[trim={5mm 5mm 5mm 1mm},clip,width=0.8\textwidth]{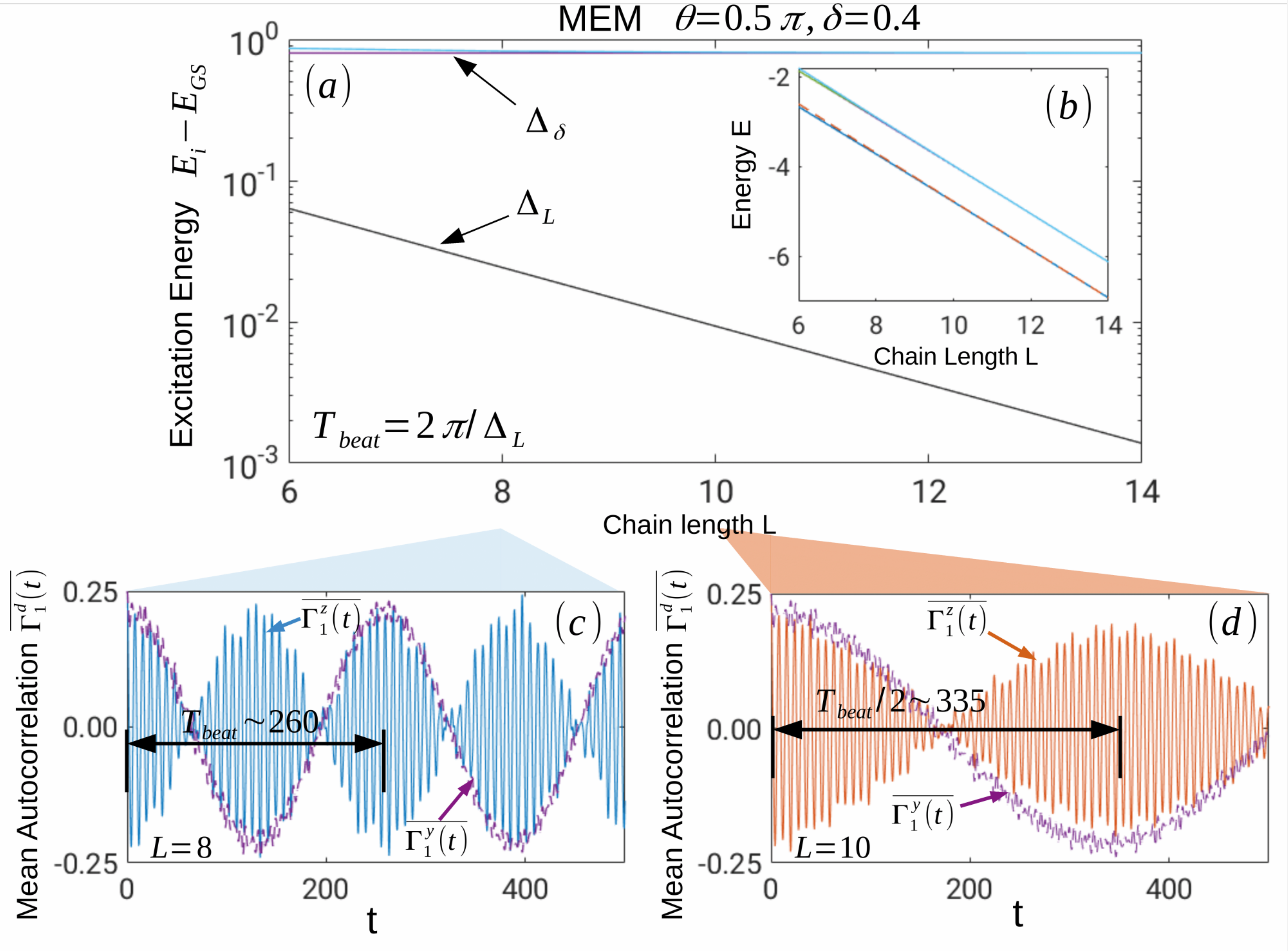}
\caption{Comparison of static energy gaps and the mean autocorrelation $\overline{\Gamma^\text{D}_\text{i}(t)}$ at an edge site, $i=1$, for a systems of varying size $L$. 
(a) Log-plot of the finite-size excitation energies, showing a bond-breaking YY-chain energy gap, $\Delta_\delta=2\delta$, and the zero-mode gap, $\Delta_\text{L}=E_1-E_\text{GS}$, as functions of system size $L$. We note that $\Delta_\delta$ is independent of system size $L$, and that $\Delta_\text{L}$ is repeated in the entire spectrum.
The inset (b) shows the bare energy of the lowest four levels for various system sizes L. The two characteristic frequencies in the temporal evolution of $\overline{\Gamma^\text{D}_\text{i}(t)}$ correspond to the transitions $E_\text{GS} \leftrightarrow E_3$ and $E_{1} \leftrightarrow E_2$. 
(c) and (d) show the long-time evolution of the edge autocorrelation function Eq.~\ref{autocorrelation} for chains of length $L=8$ and $L=10$, respectively. The characteristic frequency $\propto \Delta_\delta$ does not change with $L$. The revivals of the oscillations correspond to a beat frequency $f_\text{beat} \propto \Delta_\text{L}$ which is, however, directly dependent on the zero mode gap, $\Delta_\text{L}$, which in turn depends on $L$. The expected beat frequencies calculated directly from the spectral gaps are $T
^\text{L=8}_\text{beat}= 259$ and $T^\text{L=10}_\text{beat}= 673$, which is in good agreement with the observed dynamics.}
\label{Fig4}
\end{figure*}

Comparing the evolution of different spin-components in Fig.~\ref{Fig2b} a we see that $\sigma^\text{x}_1$ has a spin precession which does not decohere, in contrast to  $\sigma^\text{z}_1$. We note that the $x$-component of the edge spin
commutes with $\hat{H}_\text{KH}(\theta=\pi/2)$, and using the same technique as for $\sigma^\text{z}_1$ one may again calculate the spin precession of $\sigma^\text{x}_1$ from the perturbation $\hat{V}^{\text{Ising}}_\delta$. This gives a time dependence $\propto (\cos^2{\delta t} - \sin^2{\delta t})$ without any decoherence, owing to the fact that $\sigma^\text{x}_1$ does not couple  the different zero mode eigenstates.
We find that the  spin precession frequency of $\sigma^\text{x}_1$ and $\sigma^\text{z}_1$ is independent of the chosen initial state, since it explicitly depends on global gaps $\Delta_\delta$ in the spectra.  These correspond to the gap within the edge unit cell, given by the perturbation $\delta\hat{S}^\text{y}_{\text{i}}S^\text{y}_{\text{i+1}}$.
Since $\sigma^\text{x}_1$ toggles between levels within the unit cell split by $\delta$, but not between zero mode eigenstates, it only corresponds to the oscillation frequency from the gap $\Delta_\delta$.
The operator  $\sigma^\text{z}_1$ on the other hand, toggles between \textit{both} zero modes and unit cell energy levels. 
The decoherence time now relates directly to the toggling between different zero modes, corresponding to gaps $\Delta_\text{L}$ given by the commutation of the zero modes and the Hamiltonian. This correspondence is shown in Fig.~\ref{Fig4}, where the autocorrelation is evaluated for longer times.
Away from the Kitaev point $\theta_\text{KP}$ the global gaps are no longer present, so the oscillations decay quickly in both the TLL and XY phase.

\begin{figure*}
\includegraphics[width=0.8\textwidth]{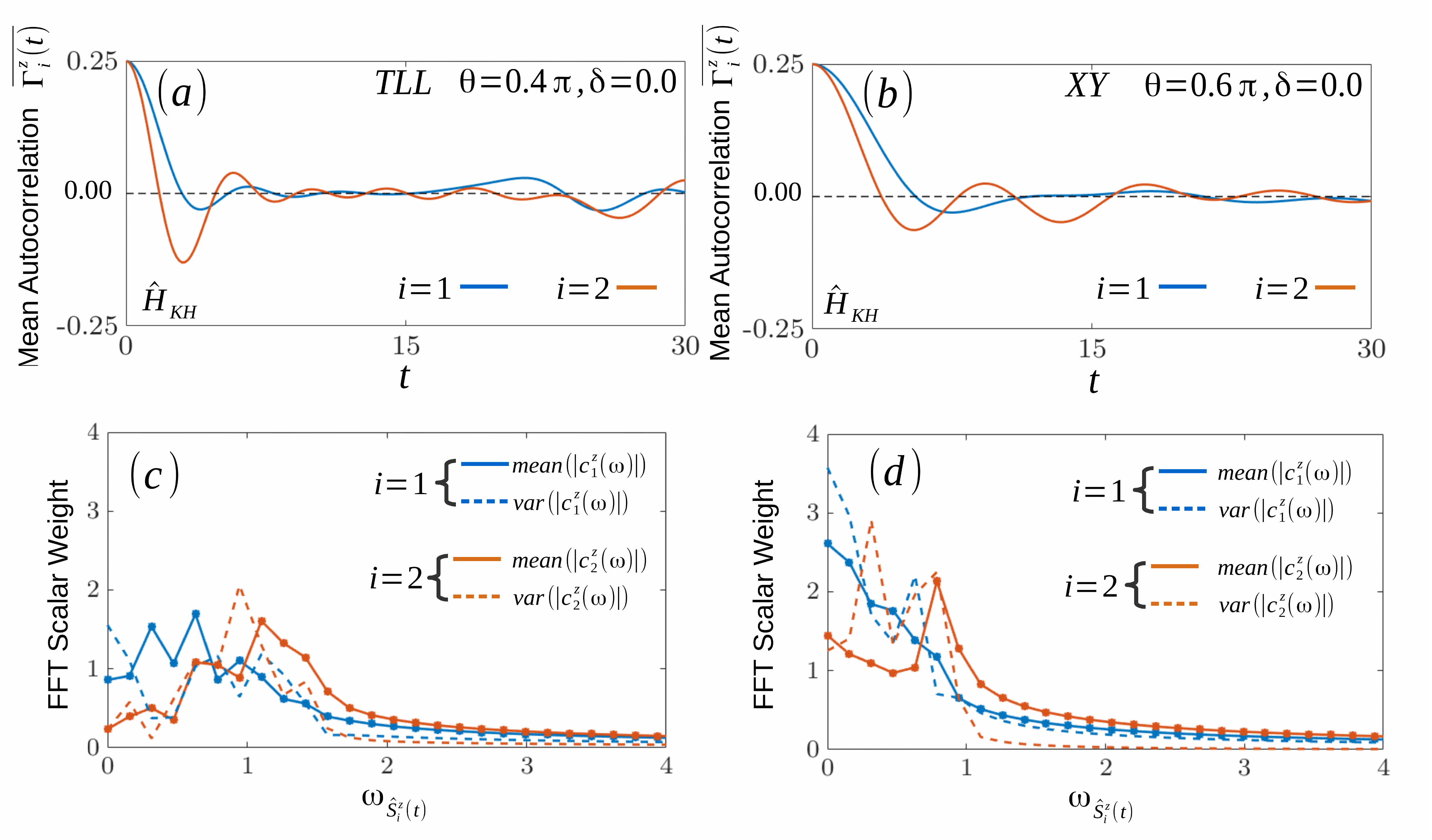}
\caption{Same as Fig.~\ref{Fig2b}, 
but for the TTL phase (a and c) and the XY phase (b and d). Each of these phases displays a broad distribution of frequencies, with high variance both in the bulk and at the edge. This is manifested in the rapid decay of the real-time oscillations in (a) and (b). This is in strong contrast to the edge oscillations in the MEM phase, providing a clear signal for the onset of Majorana edge modes at $\theta=0.5\pi, \delta>0$.}
\label{Fig3} 
\end{figure*}

Fig.~\ref{Fig3} shows that the persistent edge oscillations are entirely absent in the other parts of the phase diagram, where zero modes are not present.
In these cases there are no global degeneracies in the Hamiltonian, and oscillations at the edges simply decay in the same manner as for spins in the bulk, as seen from Fig.~\ref{Fig3}.
Simulations show that the decoherence becomes more profound, meaning that the system is fully decohered for a longer time without signs of revival,
as the systems grow in size. 
We stress that the edges are still interacting with the bulk in this MEM phase, which becomes apparent from the fact that the coherence times increase with $L$ (not shown here).
This means that we cannot think of the edges and bulk as two completely separate systems described in terms of a tensor product between them. However, observables like $\Gamma^\text{x}_{1}(t)$ may still have trivial behavior if they fully commute with the Hamiltonian, as seen in Fig.~\ref{Fig2b}.

From the above results, we conclude that the autocorrelation function can reveal a clear signature for the existence of zero modes in the Hamiltonian, providing the correct spins are measured. 

The oscillations are only visible for the edges and not in the bulk since the bulk spins generally decohere fast~\cite{fendley2016strong}.
A handy explanation is offered in the limit $\delta\gg 1$ where we notice that applying a local spin-$z$ at the edge adds the energy cost of breaking (or creating) exactly one antiferromagnetic bond: 
The operator $\hat{\sigma}_1^\text{z}$, which acts as a spin-flip operator on a local spin-$y$ state at the edge, will necessarily break \textit{or} create exactly one bond, connecting states separated in energy by the gap $\Delta_\delta=\delta$. These gaps are present throughout the spectrum and result in a large set of coherent terms in the autocorrelation function~\ref{autocorr_bare}, giving a significant contribution for long times. 
For $\hat{\sigma}_{i}^\text{z}$, acting on a bulk state $L>i>1$, a spin flip is instead associated with either simultaneous creation \textit{and} destruction of one bond (or simultaneous creation or destruction of \textit{two} bonds) and will thus connect states with energy differences smaller than the gap.
This gives rise to a set of low frequencies contributing to the temporal evolution of the bulk spin, causing an effective decay of the oscillations. 
In the limit of small $\delta$, this discrepancy between bulk and edge is reduced since $\Delta_\delta$ is then of the same order as other, small, gaps in the spectrum.
From the dynamical simulations, we find that the number of dominating frequencies for the first bulk spin corresponds to $N_\text{peaks}=L-2$ which means that for successively larger systems, the oscillations in real time will be washed out for this site. The high relative variance for the bulk spins, see Fig.~\ref{Fig2b} furthermore shows that the exact dynamics here depend heavily on the input state.
Directly at the Kitaev point, the number of dominating frequencies corresponds to the system size $L$ ($N_\text{peaks}=L/2$). The frequencies are independent of the site index and input state, whereas their relative weights depend on the site index. This explains why no coherent oscillations are seen directly at the Kitaev point.

In the limit of long-time spin precession for the MEM phase we observe decay and revival of the oscillation at the edge, resulting in a beating pattern.
The beating pattern, with period $T^\text{L}_\text{beat}$, observed in Fig.~\ref{Fig4} is a finite-size effect directly related to the revival of decohering spins, which was noted for the pure Ising model in Refs.~\cite{kemp2017long,fendley2014free}, where the revival time scaled with system size.
The envelope function in Eq.~\ref{beating} is given by the same function as for the spin $\sigma^\text{y}_1$, which has some important consequences. We know that the decoherence and revival times of the autocorrelation function here are related to the correction term $\mathcal{C}$~\cite{kemp2017long}, so these properties are therefore identical for $\sigma^\text{y}_1$ and $\sigma^\text{z}_1$. The revival time scales with the few-body gap $\sim 1/\Delta_\text{L}$, as evident in Fig.~\ref{Fig4}.
As noted before, $\Delta_\text{L}$ is identical for $\hat{H}^{\text{Ising}}_{\delta}$ and $\hat{H}^{\text{inter}}_{\delta}$. We have shown that the long-time properties of these two systems can be mapped onto those of an effective Ising model with two-site unit cells. The complete revival and resulting beating pattern in the autocorrelation plots of Fig.~\ref{Fig4}, scaling with $T^\text{L}_{\text{beat}} \propto 1/\Delta_\text{L}$, is therefore not surprising.


\section{Quantum simulation with trapped ions\label{sect-quantumsimulation}}
For a possible quantum simulation of the MEM phase, we here sketch a setup with ions trapped utilizing radio frequency (RF) fields~\cite{haffner2008quantum,wineland1998experimental,leibfried2003quantum}. 
Experiments with such systems typically realize effective Ising, XY, or XYZ spin-spin interactions, and may be supplemented with global transverse fields terms~\cite{porras2004effective,deng2005effective}.
These setups have been used in a large number of studies, for example simulation of quantum magnets~\cite{friedenauer2008simulating} entanglement propagation~\cite{jurcevic2014quasiparticle}, and variations on the quantum Ising spin chains~\cite{zhang2017observation,pagano2020quantum,liu2019confined}, along with more general quantum computing implementations~\cite{kielpinski2002architecture,gulde2003implementation}.
\begin{figure*}
\includegraphics[trim={5mm 5mm 5mm 1mm},clip,width=0.55\textwidth]{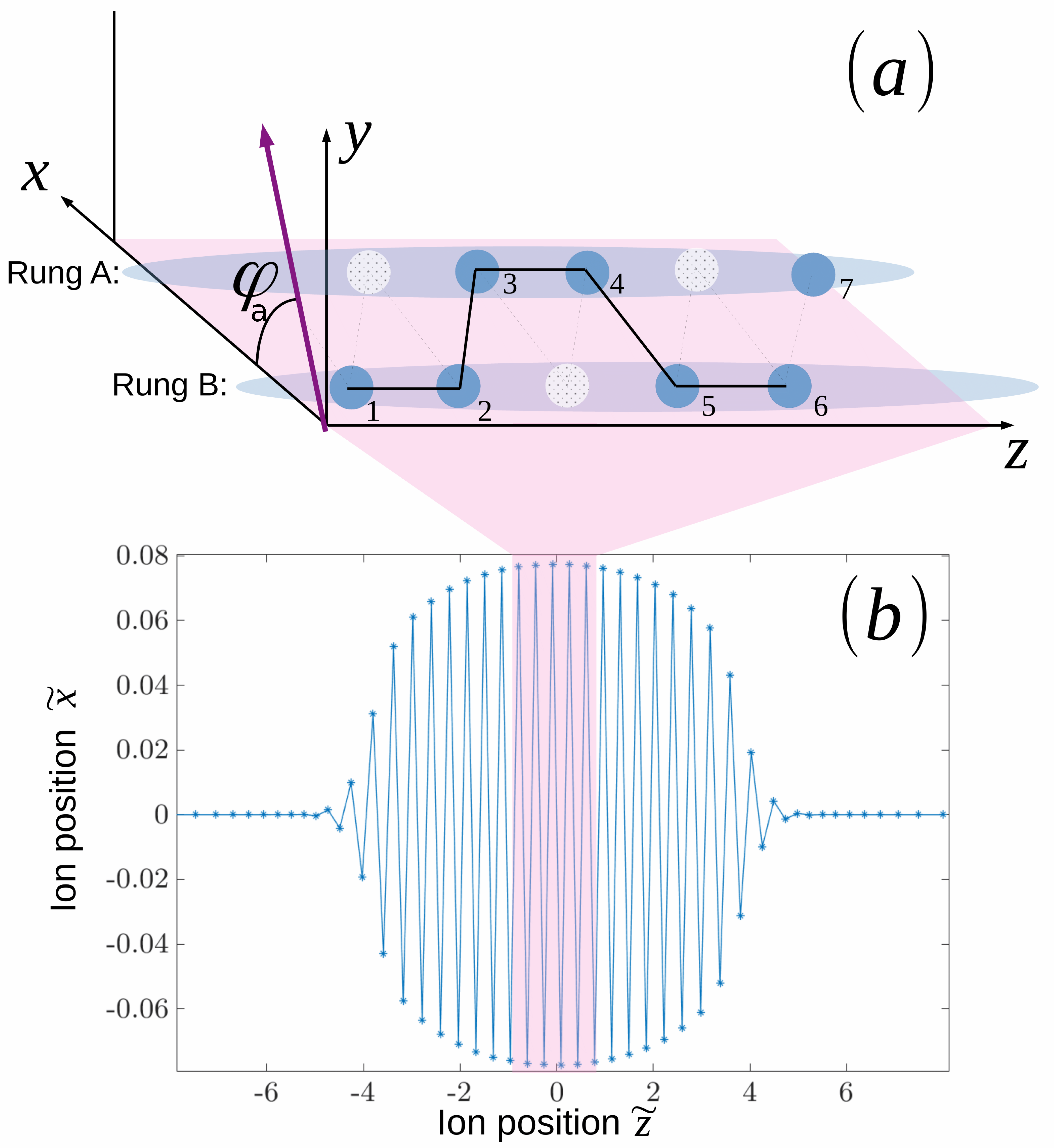}
\caption{(a) Sketch of the zig-zag configuration of ions in an RF trapping potential. The direction 
of optical beams are given by $\varphi_a \sim \pi/2$ in the $xy$-plane. Every third ion is hidden, as indicated by the white circles, producing the effective order of interacting two-level systems denoted by the numerals.
(b) Numerically simulated equilibrium positions of $N=70$ ions for the considered trap parameters $\omega_\text{x}=18.75\omega_\text{z}$ and $\omega_\text{y}=125\omega_\text{z}$. We assume that all ions outside the central region are hidden. so that the distance between different rungs is roughly uniform for the participating ions. Alternatively, a similar uniform spacing can be achieved by shaping the trapping potential~\cite{lin2009large}.}
\label{Fig5} 
\end{figure*}
Whereas most quantum simulation experiments with trapped ions use a linear configuration we will consider a geometry where  the trapping frequencies perpendicular to the tap axis are very different. For suitable parameters this ensures that the ions form a planar zig-zag structure as shown in Fig.~\ref{Fig5}. 
The use of a zig-zag configuration allows for the direction and magnitude of a laser field to control the size and sign of effective interactions between different rungs of the zig-zag spin ladder~\cite{bermudez2012quantum} and we will exploit this below. Furthermore, we assume that every third ion in our setup is selectively hidden, requiring individual addressing of ions~\cite{naegerl1999laser}, so that they do not participate in the simulation. The resulting pattern is sketched in Fig.~\ref{Fig5}.
To ensure a uniform distance between ions, we only consider the central part of a crystal containing  $N\sim100$ ions and assume that all other ions do not participate in the simulation, e.g. because they have been optically pumped to a different internal level~\cite{naegerl1999laser}, see Fig.~\ref{Fig5}. Alternatively, a more uniform distribution could also be obtained by carefully adjusting the local trapping potential~\cite{lin2009large}.

For convenience, we first employ a simple rotation on the Hamiltonian in Eq.~\ref{Hamiltonian} so that
\begin{equation}
    X\rightarrow Z_\text{sim} \quad Y\rightarrow X_\text{sim} \quad Z\rightarrow Y_\text{sim}
    \label{rotation}
\end{equation}
where the subscript "sim" denotes the axes of the simulated Hamiltonian.
We aim to realize the Hamiltonian 
\begin{equation}
\begin{split}
\hat{H}&=\sum^{\text{L/2}}_{\text{j'=1}}  V_\text{j'}^\text{a,odd} \hat{S}^\text{z}_\text{2j'-1} \hat{S}^\text{z}_\text{2j'}\\
    & +\sum^{\text{L/2}}_{\text{j'=1}} V_\text{j'}^\text{b,odd} \hat{S}^\text{x}_\text{2j'-1} \hat{S}^\text{x}_\text{2j'}
    \\
    & +\sum^{\text{L/2}}_{\text{j'=1}} V_\text{j'}^\text{b,even} \hat{S}^\text{x}_\text{2j'} \hat{S}^\text{x}_{2j'+1}
    +\text{H.C.}
\label{hamiltonian_sim1}
\end{split}
\end{equation}
with the purpose of simulating the effective Hamiltonian $\hat{H}^{\text{Ising}}_\delta=\hat{H}_\text{KH}(\theta=\pi/2)+ \hat{V}^{\text{Ising}}_\delta$ from the previous section.
We note that the indices $j'$ correspond to the effective indicies in the simulated Hamiltonian, which correspond to the \textit{active} (not hidden) ions inside the ion trap, as shown in Fig.~\ref{Fig5} a. 
For the choice $V_\text{j'}^\text{a,odd}=V_\text{j'}^\text{b,odd}=V_\text{j'}^\text{b,even}/2$ the Hamiltonian $\hat{H}_\text{sim}$ reduces to $\hat{H}_{\delta_{\text{eff}} }^{\text{Ising}}$ with $K=1$ and $\delta_\text{eff}=1$, assuming the couplings $V_\text{j}^\text{a,odd}$,$V_\text{j}^\text{b,odd}$ and $V_\text{j}^\text{b,even}$ are uniform. 

As we discuss below a major challenge is to remove additional interactions induces by the coupling mechanism corresponding to  next-nearest neighbor (or higher) interactions in each of the effective ZZ and XX interactions. 
The residual interactions will cause decoherence and need to be reduced enough for the beating mechanism to be observed at different system sizes $L$. We find that a suitable parameter regime to aim for is $\delta_\text{eff} \sim 1$.

The signature of the MEM phase of $\hat{H}_\delta^{\text{Ising}}$ requires the edge spin to be initially prepared in an eigenstate of $\hat{S}_1^\text{y}$. This can be achieved by optical pumping and subsequent $\pm \pi/2$ rotations around the x-axis~\cite{olmschenk2007manipulation,hayes2010entanglement}.
The remaining spins can be prepared into any mixed state, but to be consistent with the previous sections we here assume their initial states to also be eigenstates of $\hat{S}^\text{y}_\text{i}$.
With the initial spin-state $s^\text{y}_1$ known, the autocorrelation function ${\Gamma^\text{y}_\text{i}(t)}$
 in Eq.~\ref{autocorrelation} can be evaluated by measuring the spin along $y$ at a later time $t$.
 
We now proceed to discuss the realization of the two different interactions in $\hat{H}_\text{sim}$, starting with the ZZ interaction.
For a detailed derivation and discussion of parameters of the ion-trap simulation, see the Supplemental materials~\ref{asect-quantumsim}.

\subsection{Simulating ZZ and XX interactions}
For the ZZ interaction we consider a two-photon $\Lambda$ scheme where the lasers coupling two stable ground levels $|\uparrow\rangle$ $|\downarrow\rangle$ to an excited state are far-off resonant with the dipole allowed transition and the detuning is given by $\mathcal{D}$ which is much larger than the spontaneous decay rate of the system.
In line with Refs.~\cite{bermudez2012quantum,porras2004effective} we employ a pair of laser fields with effective Rabi frequencies $\Omega_{1,\varsigma}$ and $\Omega_{2,\varsigma}$ for $\varsigma=\uparrow, \downarrow$ coupling the ground levels $\varsigma$ to the excited state. 
The effective Raman wavevector of the two fields is given by $\vec{k}^\text{a}_{l}=\vec{k}^\text{1}-\vec{k}^\text{2}
=k^\text{a}_{l}\{ \cos{\varphi_\text{a}}, \sin{\varphi_\text{a}}, 0 \}$, and can be tuned via alignment of the lasers.
The laser beatnote $\omega^\text{a}_l=\omega_1-\omega_2$ of the fields is chosen close to the ions' collective vibrational motion in the transversal direction $y$, with mode energies $\omega^\text{y}_\text{p}$, and far-off resonance with the vibrational modes in the zig-zag plane with mode energies $\omega^\text{xz}_\text{p}$. The transverse vibrational modes act as mediators of an effective spin-spin interaction of the canonically transformed Hamiltonian~\cite{porras2004effective}, and 
by carefully choosing the detunings and alignments of the laser fields a non-isotropic effective ZZ interaction with tunable strength and range can be realized~\cite{bermudez2012quantum}. 
\begin{equation}
\begin{split}
    \tilde{V}^\text{a}_\text{ij}= \sum_{\text{ij}} V^a_\text{ij} \hat{\sigma}^z_\text{i}\hat{\sigma}^z_\text{j}
\end{split}
\end{equation}
The overall strength of the interaction is controlled by the magnitude and direction of the laser fields, affording some freedom in choosing the parameters in our effective Hamiltonian.
Crucially, the factor $ V^a_\text{ij}$ is dependent the alignment of the field and the relative equilibrium positions $\vec{\tilde{r}}^0_\text{ij}$ of the interacting ions, so that the non-homogeneity of the interaction can be tailored via the fields.
\begin{equation}
\begin{split}
    {V}^\text{a}_\text{ij}\propto& - \Omega_\text{a}^2 (\sin{\varphi_\text{a}})^2\\
    &\cos{ \left(k^a_l ( x_\text{j}-x_\text{i} )\cos{\varphi_\text{a}} +( y_\text{j}-y_\text{i} )\sin{\varphi_\text{a}} \right)} 
\end{split}
\end{equation}
The effective two-photon Rabi frequency is given by $\Omega_\text{a}=\left(\Omega_{1,\downarrow}\Omega_{2,\downarrow}^*+\Omega_{1,\uparrow}\Omega_{2,\uparrow}^* \right)/2\mathcal{D}$.

For the zig-zag configuration of the ions, it is convenient to choose $\vec{k}^\text{a}_{l}$ in the $xy$-plane, i.e. perpendicular to the direction of the rungs, as shown in Fig.~\ref{Fig5} so that
\begin{equation*}
\begin{split}
 &\text{Same rung:} \\
 &
 \cos{\left( \vec{k}^\text{a}_{l} \cdot \tilde{\vec{r}}^0_\text{ij}\right)} 
 =1 \\
  &\text{Different rungs:}
  \\
  &
  \cos{\left( \vec{k}^\text{a}_{l} \cdot \tilde{\vec{r}}^0_\text{ij}\right)} 
  =
   \cos{ \left( \left( {x}_\text{i}-x_\text{i+1}\right)\cdot\cos{\varphi_\text{a}}\right)}
\end{split}  
\end{equation*}
By choosing the angle $\varphi_a$ we can now eliminate the interaction between \textit{different} rungs, even when the optical wavelength is small relative to the mutual ion distances~\cite{bermudez2012quantum}. 

We numerically calculate the equilibrium configuration of the ion trap with $N=70$ to find the ions' positions and their transverse vibrational eigenmodes.
We employ these quantities to evaluate the full expression for the effective interaction given in the Supplemental materials~\ref{asect-quantumsim}. 
Choosing the detunings of the fields, relative to the vibrational mode energies along $y$ such that $\left(\omega^\text{y}_\text{p}- \omega^\text{a}_{l}\right)/\omega^\text{y}_\text{p}\sim 0.05$, we find that the typical distance dependence of the interaction strength  becomes $\sim 1/|i-j|^\text{R}$ with $R\sim 3$. Since this falls off of quickly with the distance, the interaction will be dominated by NN interactions~\cite{monroe2021programmable}. 
We see from Fig.~\ref{Fig5} a that by hiding every third ion, we can map the same rung NN interaction to odd indices in the effective system 
whereas different rungs corresponds to NN interactions starting on even indices, i.e.
\begin{equation}
\begin{split}
    \text{Different rung:} \quad &\tilde{V}^\text{a}_\text{i,i+1}\rightarrow V^\text{a}_\text{2i',2i'+1}\\
    \text{Same rung:} \quad&\tilde{V}^\text{a}_\text{i,i+2}\rightarrow V^\text{a}_\text{2i'-1,2i'}
    \label{mapping}
\end{split}
\end{equation}
for the effective indices $i'$ of the active ions.

For the effective XX interaction we need to drive a transition between two internal levels. This can either be done directly or as a two-photon Raman transition. The effective spin coupling is implemented via the vibrational sidebands of the transition~\cite{sorensen1999quantum}. 
The angular frequencies of the driving are given by $\omega^\text{b}_1=\omega^\text{b}-\omega^\text{b}_{l}$ and $\omega^\text{b}_2=\omega^\text{b}+\omega^\text{b}_{l}$. Here $\omega^\text{b}$ is the transiton frequency between the considered internal levels of the ions. The detuning $\omega^\text{b}_{l}$ is roughly matched to the transverse trapping frequency $\omega_\text{y}$ so that $\omega^\text{b}_{l}\approx\omega_\text{y}$. We aim to virtually excite the vibrational sidebands, and we employ the sideband detuning $\gamma^\text{b}=\omega^\text{b}_{l}-\omega_\text{y}$. The sideband detuning is chosen to be positive, so that the $\omega^\text{b}_{l}$ lies above the highest vibrational mode along $y$ (out of plane), which is the centre of mass mode $\omega_\text{c.o.m.}$. 
We employ the same vibrational branch as for ZZ, but we assume there are no interference effects between the processes implementing ZZ and XX. This can be achieved by ensuring that the frequencies are incommensurate.  
The virtual phonon exchange between ions induces an effective interaction
\begin{equation}
    \tilde{V}^{b}_\text{ij}= |\Omega_{b}|^2 \frac{\hbar |k^\text{b}_{l}|^2}{4} \sum_\text{p} \frac{M_\text{i,p} M_\text{j,p}}{ \left(\omega^\text{b}_{l}\right)^2  - \omega_\text{p}^2} 
    \label{XX_int}
\end{equation}
where the vibrational mode eigenvectors $M_p$ and mode energies $\omega_p$ are for  the strongly confined direction y~\cite{molmer1999multiparticle}. 

We again employ the calculated vibrational modes and ion positions to explicitly calculate the effective interaction. The overall strength of the XX interaction can be controlled by the effective Rabi frequency $\Omega_\text{b}$~\cite{sorensen1999quantum}.
Choosing the detuning $\left(\omega^\text{y}_\text{p}- \omega^\text{b}_{l}\right)/\omega^\text{y}_\text{p}\sim 0.05$, we may realize approximate interactions $\sim 1/|i-j|^\text{R}$ with $R\sim 3$ for the XX interaction, so that residual terms are of similar order as for ZZ. The detuning is chosen to be different than that for the ZZ interaction, but on the same order of magnitude.
Since there is no angular factor in Eq.~\ref{XX_int} the effective interaction connects both sites within the same rung and sites between rungs.

Putting everything together, we obtain for the zig-zag indicies $i,j$
\begin{equation}
\begin{split}
\hat{\tilde{H}}_\text{sim}&=\sum^{\text{L}}_{\text{i}} \left( \tilde{V}_\text{i,i+2}^\text{a} \hat{S}^\text{z}_\text{i} 
\hat{S}^\text{z}_\text{i+2}
+ \tilde{V}_\text{i,i+2}^\text{b} \hat{S}^\text{x}_\text{i} 
\hat{S}^\text{x}_\text{i+2}\right)\\
&+\sum^{\text{L}}_{\text{i}}  \tilde{V}_\text{i,i+1}^\text{b} \hat{S}^\text{x}_\text{i} 
\hat{S}^\text{x}_\text{i+1}\\
    &+\sum^{\text{L}}_{|i-j|>2} \left(\tilde{R}^\text{a}_\text{ij}\hat{S}^\text{z}_\text{i} \hat{S}^\text{z}_\text{j}+ \tilde{R}^\text{b}_\text{ij}\hat{S}^\text{x}_\text{i} \hat{S}^\text{x}_\text{j} \right)+ \text{H.C.}
\label{hamiltonian_sim2}
\end{split}
\end{equation}
We now use the mapping in Eq.~\ref{mapping} to transform into the indices $i'j'$ for the \textit{active} ions
\begin{equation}
\begin{split}
\hat{{H}}_\text{sim}&=\sum^{\text{L/2}}_{\text{i'}} \left( {V}_\text{2i'-1,2i'}^\text{a} \hat{S}^\text{z}_\text{2i'-1} 
\hat{S}^\text{z}_\text{2i'}
+ {V}_\text{2i'-1,2i'}^\text{b} \hat{S}^\text{x}_\text{2i'-1} 
\hat{S}^\text{x}_\text{2i'}\right)\\
&+\sum^{\text{L/2}}_{\text{i}}  {V}_\text{2i',2i'+1}^\text{b} \hat{S}^\text{x}_\text{2i'} 
\hat{S}^\text{x}_\text{2i'+1}\\
    &+\sum^{\text{L}}_{|i'-j'|>1} \left({R}^\text{a}_\text{i'j'}\hat{S}^\text{z}_\text{i'} \hat{S}^\text{z}_\text{j'}+ {R}^\text{b}_\text{i'j'}\hat{S}^\text{x}_\text{i'} \hat{S}^\text{x}_\text{j'} \right)+ \text{H.C.}
\label{hamiltonian_sim_done}
\end{split}
\end{equation}
This form of the Hamiltonian agrees with the desired model in Eq.~\ref{hamiltonian_sim1} apart from the residuals in the last line. According to the arguments above these residuals can be rather small. Furthermore, for a translationally invariant system, the desired coefficients will be identical between units cells as desired and can be adjusted to the desired values ratio to realize the MEM phase of the Kitaev-Heisenberg model.

For a real ion trap, the parameters can suffer from numerous imperfections. In particular, for a standard ion trap, the density will be higher near the center of the trap and the ions will not be equidistant, as shown in Fig.~\ref{Fig5}. In principle this can be overcome by carefully designing the trapping potential~\cite{pagano2018cryogenic,johanning2016isospaced}, but below we explore the limitations imposed by operating in a standard ion trap with harmonic confinement. 

To investigate the imperfections in a real ion trap we consider the situation depicted in Fig.~\ref{Fig5}, consisting of 70 trapped ions, see the Supplemental materials~\ref{asect-quantumsim} for further details. 
To evaluate the role of imperfections the coefficients $V_\text{ij}^\text{a}$,$V_\text{ij}^\text{b}$ and residuals are evaluated for the numerically calculated ion positions and eigenstates. 
We specifically do this by first fixing $|k^\text{a}_{l}|$,$|k^\text{b}_{l}|$ and then the angle $\varphi_a$ so that different rung interactions disappear for ZZ, and proceed to choose the exact detunings for all fields.
We can then finally match the Rabi frequencies $\Omega_\text{a}$ and $\Omega_\text{a}$ so that  
${V}_\text{2i',2i'+1}^\text{a} \sim {V}_\text{2i',2i'+1}^\text{b} \sim {V}_\text{2i',2i'+1}^\text{b}/2$, mapping the simulated Hamiltonian~\ref{hamiltonian_sim_done} onto $\hat{H}^{\text{Ising}}_\delta=\hat{H}_\text{KH}(\theta=\pi/2)+ \hat{V}^{\text{Ising}}_\delta$ with $\delta_\text{eff}\sim 1$, with additional residual terms.
Using the numerically coupling coefficients we can then proceed to evaluate the dynamics in the MEM phase corresponding to the time evolution in Fig.~\ref{Fig4}. 

\begin{figure*}
   \includegraphics[trim={3mm 3mm 3mm 3mm},clip,width=0.7\textwidth]{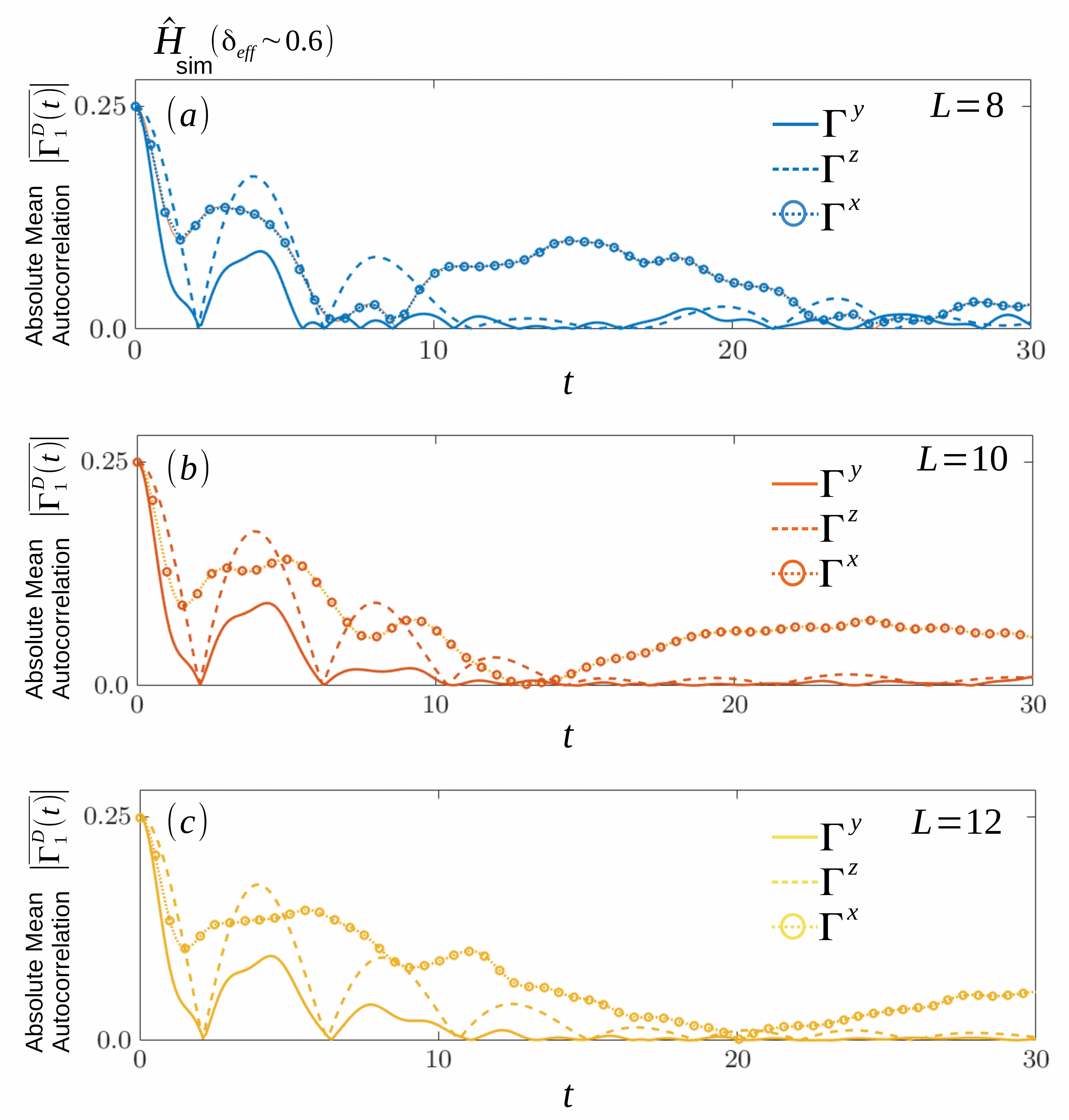}
\caption{Time evolution of the absolute mean autocorrelation $|\overline{\Gamma^\text{D}_\text{1}(t)}|$ of an the at an edge site, $i=1$ from a numerical simulation of the ion trap Hamiltonian in Eq.~\ref{hamiltonian_sim_done}.
For the parameters considered in the Supplemental materials~\ref{asect-quantumsim}, we realize $K=1, \delta_\text{eff}\sim0.61$ with residuals $\mathcal{O}(10^{-1})$. In principle other values of $\delta_\text{eff}$ can be simulated by adjusting the setup. 
The set of initial states are chosen as pure eigenstates to the operator $\hat{S}^\text{D}_i$. 
Plots (a)-(c) indicates that by increasing the number  of active ions in the simulation, one can observe a heavily damped beating pattern in $|\overline{\Gamma^\text{y}_\text{1}(t)}|$, corresponding to the undamped beating pattern observed for $|\overline{\Gamma^\text{z}_\text{1}(t)}|$ in Fig.~\ref{Fig2b} a (for the idealized Hamiltonian). Note that the simulated Hamiltonian is rotated relative to the idealized Hamiltonians via Eq.~\ref{rotation}, so that  the oscillation for $|\overline{\Gamma^\text{y}_\text{1}(t)}|$  indeed corresponds to the oscillation  for $|\overline{\Gamma^\text{z}_\text{1}(t)}|$ in Fig.~\ref{Fig2b} a. 
We see that $|\overline{\Gamma^\text{x}_\text{1}(t)}|$ exhibits revivals of the long-time coherence scaling with $L$. This is due to the presence of a strong zero mode in the Hamiltonian.
The coherence time $t^\text{z}_\text{c}$ of the oscillation in $|\overline{\Gamma^\text{z}_\text{1}(t)}|$, on the other hand, is more independent of system size since it does not correspond to a zero mode. Its decoherence is explained by the residual terms in the quantum simulation and gives an overall estimate of the influence of impurities in the simulation.   
The oscillation of $|\overline{\Gamma^\text{y}_\text{1}(t)}|$  
should exhibit a beating pattern in the limit of an idealized Hamiltonian, but decoheres on the same time-scale as  $|\overline{\Gamma^\text{z}_\text{1}(t)}|$. There is however an indication of the first node in the beating in $|\overline{\Gamma^\text{y}_\text{1}(t)}|$ for $L=8$, which occurs before the system has fully decohered. This is seen from the node in $|\overline{\Gamma^\text{y}_\text{1}(t)}|$ at $t\approx 6$, which is pushed towards longer times for larger system sizes $L$.
}
\label{Fig6} 
\end{figure*}

The results of a numerical simulation, performed for $\sim40$ sampled initial states, of the ion trap are shown in Fig.~\ref{Fig6}. 
For the simulated setup the residual terms, the largest of which correspond to $\sim 10 \%$ of ${V}^\text{a}$ and ${V}^\text{b}$, clearly have a large effect on the results. 
The beating pattern visible for the idealized Hamiltonian in Eq.~\ref{hamiltonian_sim1} is hard to observe due to the rapid decay of oscillations in the autocorrelations. We can however see the size-dependent revivals for the long-time coherent spin along one of the axes, here along $x$ (Note that when comparing to the idealized Hamiltonian we use the rotation~\ref{rotation}).
In Fig.~\ref{Fig4}, showing the dynamics of the idealized Hamiltonian, we further had two precessing spin components which were enveloped by a long-time beating. 

Here we also saw that the decoherence of one of the precessing spin components ($z$) was enveloped by the long-time coherent autocorrelation function for the $y$-compoenent, while the other precessing spin component  ($x$), evolves more independently from the long-time coherent spin. 
This is behavior is somewhat visible also in the ion-trap simulation when comparing Fig.~\ref{Fig6} a and Fig.~\ref{Fig6} b. We see that the oscillation of $|\overline{\Gamma^\text{z}_\text{1}(t)}|$, which does not correspond to a zero mode, is less dependent on system size than the other spins. We conclude this by noticing that the oscillation of $|\overline{\Gamma^\text{z}_\text{1}(t)}|$ is not suppressed in the region around the first node in $|\overline{\Gamma^\text{x}_\text{1}(t)}|$ for $L=8$.
However, the amplitude of the oscillating $|\overline{\Gamma^\text{y}_\text{1}(t)}|$ is suppressed in this region.
For $L=10$, the oscillation in $|\overline{\Gamma^\text{y}_\text{1}(t)}|$ is instead suppressed at slightly later times. This is due to the longer coherence time of $|\overline{\Gamma^\text{x}_\text{1}(t)}|$, which envelopes the oscillation, as shown in section~\ref{asec_spinbeating}.
It is however evident that the precession in $|\overline{\Gamma^\text{y}_\text{1}(t)}|$ also suffers from the same type of decoherence as $|\overline{\Gamma^\text{z}_\text{1}(t)}|$, since there is no visible revival for $|\overline{\Gamma^\text{y}_\text{1}(t)}|$. This gives an estimate for the influence of the residual interactions in the simulated Hamiltoninan, since $|\overline{\Gamma^\text{z}_\text{1}(t)}|$ should not decohere in the limit of zero residuals. (Compare to the evolution of the corresponding $|\overline{\Gamma^\text{x}_\text{1}(t)}|$ of the idealized Hamiltonian in Fig.~\ref{Fig2b} a )
The behavior of the spins away from the edge (not shown) is similar for the simulated Hamiltonian as for the idealized Hamiltonian, where the long time coherence is seen also for $s_2$ but not for $s_3$. There is also no spin precession away from the edge for either spin.

In the ideal case, the nodes in the beating and long-time coherent spin autocorrelation can be used as a direct measure of the finite-size Majorana gap $\Delta_L=E_1-E_{GS}$, which effectively simulates finite-size scaling in the MEM phase. The residual interactions in the quantum simulation however cause the oscillation of the edge spin to decohere rapidly, generally before the first node. Furthermore, the residual interactions destroy the global degeneracy corresponding to the zero modes, so the gaps throughout the spectrum are no longer homogeneous for the simulated Hamiltonian. 
By making a more homogeneous distance between the ions and implementing additional fields at different detunings and angles into the quantum simulation, the residuals can be reduced. This would allow for the decoherence time $t^\text{z}_\text{c}$ to be increased so that the first node, and subsequent revival of the oscillation in $|\overline{\Gamma^\text{y}_\text{1}(t)}|$ can be observed for successively larger systems.

\section{\label{sec-5Conclu} Conclusions and Outlook}
We have showcased dynamical features of the Kitaev-Heisenberg Hamiltonian, particularly focusing on the behavior around the so-called Kitaev point at which a multi-degenerate set of Majorana modes appear. 
By perturbing the model with a nearest-neighbour term $\hat{S}^y\hat{S}^y$ on even sites, we see that the model can be mapped directly onto the Majorana edge-mode (MEM) phase of Kitaev's model for p-wave paired superconductors, with one additional degree of freedom per unit site. By perturbing with a uniform Ising term  $\hat{S}^y\hat{S}^y$ between all sites, the additional degrees of freedom are lost but the system retains the Majorana edge modes.

Studying the direct time-development of local spins under this Hamiltonian we show that the MEM phase of the latter system can be identified via the precession of edge-site spins, which oscillate with two dominant frequencies. This gives rise to a beating pattern corresponding to the finite-size energy gap between the semidegenerate Majorana edge-mode states. 
The precession frequency of the spin is set by the interaction within the outermost unit cell of the system, while the beating is enveloped by the long-time coherent spin dynamics, which depends on the system size L. 
This is analogous to the long-time coherence of edge-spins in the Ising model, an effect caused by strong zero modes present in the system~\cite{kemp2017long,fendley2016strong}.
We show that zero modes affect the long-time coherence of the spins measured along two different axes in our model, while the dynamical evolution of a spin prepared along the third axis is independent of zero modes, and therefore independent of system size $L$.
These characteristics are not present in other parts of the phase diagram, or for spins in the bulk.

Crucially, this method of studying the dynamical properties does not rely on the repeated and deterministic preparation of a \textit{single} initial state. It instead only requires deterministic preparation of a single spin, while the remaining spins can be randomly distributed.

We sketch an ion-trap quantum simulation, in which the steady-state zig-zag configuration of harmonically confined ions is exploited to realize the MEM phase. We see that some finite-size scaling properties of the spin dynamics can be observed, even for a setup with relatively large residual interaction terms.
If the residuals were to be further reduced, our setup and readout mechanism could realize a quantum simulation of an interesting numerical challenge; the finite-size scaling of collective Majorana edge modes.

{\bf Acknowledgements:} 
We would like to thank Lorenzo Campos Venuti for useful discussions. 
This work was supported by the Swedish Research Council under grant number 2018-00833, the US Department of Energy under grant number DE-FG03-01ER45908 and the Danish National Research Foundation (Center of Excellence ”Hy-Q”, Grant No. DNRF139).

\bibliographystyle{apsrev4-1}
\bibliography{FEB_11}

\pagebreak
\widetext
\begin{center}
\textbf{\large Supplemental Materials: Probing Majorana Modes via Local Spin Dynamics}
\end{center}

\section{Majorana modes and spectral properties at the Kitaev point\label{asec-1spectral}}
Here, we discuss the decomposition and transformation of the Hamiltonian~\ref{Hamiltonian} referenced in section~\ref{sec-2Model} of the main text. These transformations are performed to highlight the appearance of Majorana modes in the model. 
For an intuitive understanding of the spectrum of the Hamiltonian~\ref{Hamiltonian} we may rewrite it as
\begin{equation}
\begin{split}
\hat{H}_{KH}=
    (K+J)\sum^{L/2}_{j=1} (\hat{S}^x_{2j-1} \hat{S}^x_{2j} +
    \hat{S}^y_{2j} \hat{S}^y_{2j+1})+\\
    J \sum^{L/2}_{j=1} 
    (\hat{S}^y_{2j-1} \hat{S}^y_{2j} +
    \hat{S}^x_{2j} \hat{S}^x_{2j+1}).
\end{split}
\end{equation}
We can fermionize this Hamiltonian by applying the Jordan-Wigner transformation and writing
\begin{equation}
\hat{S}^x_{j} \hat{S}^x_{j+1} = \frac{(\hat{f}^\dagger_j-\hat{f}_{j})(\hat{f}^\dagger_{j+1} + \hat{f}_{j+1}) }{4},
\end{equation}
\begin{equation}
\hat{S}^y_{j} \hat{S}^y_{j+1} = - \frac{(\hat{f}^\dagger_j+\hat{f}_{j})(\hat{f}^\dagger_{j+1} - \hat{f}_{j+1}) }{4},
\end{equation}
where $\hat{f}^\dagger$ and $\hat{f}$ are the fermionic creation and annihilation operators respectively.
In the spirit of Ref.~\cite{agrapidis2018ordered} we  rewrite the sum in terms of the unit cell index $j$, where each cell contains a black(b) site and a white(w) site.
This transforms the Hamiltonian into
\begin{equation}
\hat{H}_{KH}=\frac{K+J}{4} \hat{H}_A + \frac{J}{4}\hat{H}_B,
\end{equation}
with two terms
\begin{equation}\begin{split}
    \hat{H}_A=& \sum^{L/2}_{j=1} (\hat{f}^\dagger_{b,j}-\hat{f}_{b,j})(\hat{f}^\dagger_{w,j} + \hat{f}_{w,j})\\ &-(\hat{f}^\dagger_{w,j}+\hat{f}_{w,j})(\hat{f}^\dagger_{b,j+1}-\hat{f}_{b,j+1}),
\end{split}\end{equation}
and
\begin{equation}\begin{split}
    \hat{H}_B= &\sum^{L/2}_{j=1}
    (\hat{f}^\dagger_{w,j}-\hat{f}_{w,j})(\hat{f}^\dagger_{b,j+1} + \hat{f}_{b,j+1})  \\
    &-(\hat{f}^\dagger_{b,j}+\hat{f}_{b,j})(\hat{f}^\dagger_{w,j} - \hat{f}_{w,j}).
\end{split}\end{equation}

Like in Ref. \cite{agrapidis2018ordered} we define Majorana operators as
\begin{equation}
\begin{split}
 &\hat{b}_{1,j} = \hat{f}_{b,j}^\dagger + \hat{f}_{b,j}\\
 &\hat{b}_{2,j} = i(\hat{f}_{b,j}^\dagger - \hat{f}_{b,j}) \\
 &\hat{w}_{1,j} = \hat{f}_{w,j}^\dagger + \hat{f}_{w,j} \\
 &\hat{w}_{2,j} = i(\hat{f}_{w,j}^\dagger - \hat{f}_{w,j}) \\
\end{split}
\end{equation}
We can now write
\begin{equation}\begin{split}
    \hat{H}_A=& i \sum^{L/2}_{j=1} \hat{w}_{1,j}\hat{b}_{2,j+1}-\hat{b}_{2,j}\hat{w}_{1,j}
\end{split}\end{equation}
\begin{equation}\begin{split}
    \hat{H}_B=& i \sum^{L/2}_{j=1} \hat{b}_{1,j}\hat{w}_{2,j} - \hat{w}_{2,j}\hat{b}_{1,j+1}
\end{split}\end{equation}
Finally, we define two independent non-local fermion operators
\begin{equation}
\begin{split}
 &\hat{d}^\dagger_j = \frac{\hat{w}_{1,j} - i\hat{b}_{2,j}}{2}\\
 &\hat{\tilde{d}}^\dagger_j = \frac{\hat{w}_{2,j} - i\hat{b}_{1,j}}{2}
\end{split}
\end{equation}
This allows us to finally write
\begin{equation}\begin{split}
    \hat{H}_A=&  \frac{1}{2}\sum_{j=1} \hat{d}^\dagger_j\hat{d}_j +\frac{1}{4}\sum_{j=1}(\hat{d}^\dagger_j\hat{d}_{j+1} + \hat{d}^\dagger_j\hat{d}^\dagger_{1+j}) + h.c 
\end{split} \label{fermHamA}\end{equation}

\begin{equation}\begin{split}
    \hat{H}_B=&  -\frac{1}{2}\sum_{j=1} \hat{\tilde{d}}^\dagger_j\hat{\tilde{d}}_j -\frac{1}{4}\sum_{j=1}(\hat{\tilde{d}}^\dagger_j\hat{\tilde{d}}_{j+1} + \hat{\tilde{d}}^\dagger_j\hat{\tilde{d}}^\dagger_{1+j}) + h.c 
\end{split}\label{fermHamB}\end{equation}
which describes two independent p-wave superconductors at the boundary point of the Majorana edge-mode phase.
At the Kitaev point ($\theta=\pi/2)$, only one of the Kitaev chains contributes energy in the Hamiltonian, and the system gets one free spin per unit cell, leading to degeneracies $2^{N/2}$ and $2^{N/2 -1}$ for a non-periodic and periodic chain, respectively~\cite{agrapidis2018ordered}.
Slightly tilting $\theta$ away from the Kitaev point this degeneracy is removed, but we still retain the multiplet structure.


\section{Retrieving the gapped edge mode phase in the spin-representation\label{asec_gappedmodederivation}}
Here, we construct the appropriate form of perturbations to the Kitaev-Heisenberg model to study Majorana edge modes, as discussed in section~\ref{sec-2Model} of the main text. For definitions and background consult sections~\ref{sec-2Model} and~\ref{asec-1spectral}.
Starting with the fermionic Hamiltonians~\ref{fermHamA} and~\ref{fermHamB} derived in the previous section we note that 
going away from the critical point into the edge-mode phase requires a decrease in the relative size of the local term(s) $\hat{{d}}^\dagger_i\hat{{d}}_i$ and  $\hat{\tilde{d}}^\dagger_i\hat{\tilde{d}}_i$, respectively. We therefore perturb each of the two Kitaev chains. After mapping back to the spin-representation we find that for $\theta=\pi/2$ the appropriate operator is
\begin{equation}
\begin{split}
&\hat{H}_{A,\delta} = -\delta \sum_{j=1} \hat{d}^\dagger_j \hat{d}_j \\
&\hat{H}_{B,\delta} = \delta \sum_{j=1} \hat{\tilde{d}}^\dagger_j \hat{\tilde{d}}_j 
\end{split}
\end{equation}
where index $j$ denotes the unit cell index.
We map this change back to the fermion operators.
\begin{equation}
\begin{split}
    &-\delta \sum_{j=1} \hat{d}^\dagger_j \hat{d}_j =-\frac{\delta}{4}\sum_{j=1} (\hat{w}_{1,j}-i\hat{b}_{2,j})(\hat{w}_{1,j}+i\hat{b}_{2,j})\\
    &=-\frac{\delta}{4}\sum_{j=1}(\hat{f}^\dagger_{w,j}+\hat{f}_{w,j}  +\hat{f}^\dagger_{b,j}-\hat{f}_{b,j})(\hat{f}^\dagger_{w,j}+\hat{f}_{w,j}  -\hat{f}^\dagger_{b,j}+\hat{f}_{b,j})
\end{split}
\end{equation}
Neglecting constant terms and applying the Jordan-Wigner transform we are left with
\begin{equation} \hat{H}_{A,\delta}=-\delta \sum_{j=1} \hat{{d}}^\dagger_j \hat{{d}}_j 
= -\frac{\delta}{2}\sum_{j=1}^{L/2} \hat{S}^x_{2j-1}\hat{S}^x_{2j}
\end{equation}
and similarly
\begin{equation}\hat{H}_{B,\delta}= +\delta \sum_{j=1} \hat{\tilde{d}}^\dagger_j \hat{\tilde{d}}_j =\frac{\delta}{2}\sum_{j=1}^{L/2} \hat{S}^y_{2j}\hat{S}^y_{2j+1}
\end{equation}
At the point $\theta = \pi/2$, the Hamiltonian is equivalent to $\hat{H}_{A}$. Note that the appropriate sign of the perturbation $\delta$ depends on which Kitaev point we consider.
As shown in Ref.~\cite{agrapidis2018ordered}, the absent Majorana operators map back to the spin operators
\begin{equation}
\hat{b}_{1,j}= 2 (\prod^{j-1}_{k=1} \hat{S}^z_{k',k}) \hat{S}^x_{b,j}
\label{majoranaA}
\end{equation}
\begin{equation}
\hat{w}_{2,j}= 2 \prod_{k'=w,b} (-\hat{S}^z_{k',k})(-\hat{S}^z_{b,j})( \hat{S}^y_{w,j}).
\label{majoranaB}
\end{equation}
each of which has one free index per unit cell, giving a degeneracy of the ground state.

\section{\label{asec_phases} Phases around the Kitaev point}
 We here further characterize the phases around the critical point $\delta=0, \theta=\pi/2$ for $\hat{H}_{KH}+V^\text{Ising}_\delta$. In addition to the spectral properties discussed in the main text, we here calculate the static spin structure factor
 \begin{equation}
\tilde{P}^D(q)= \frac{2}{L} \sum_{k,l} \langle \hat{S}^D_k \hat{S}^D_l \rangle e^{-iq(k-l)}     
 \end{equation} 
 where the unit cell length is taken as 1.
We also consider the von Neumann entanglement entropy of a subsystem with length $l$
\begin{equation}
\tilde{S}_L(l)=-\text{Tr}_l \rho_l \text{log}(\rho_l)     
 \end{equation} 
 with the reduced density matrix $\text{Tr}_{L-l} \rho$ for the full density matrix $\rho$.
  The relevant phases are sketched in the phase diagram in Fig.~\ref{Fig1}, and they are found to largely agree to those found in Ref.~\cite{agrapidis2018ordered}.

\begin{figure*}[h]
  \includegraphics[trim={1mm 2mm 2mm 1mm},clip,width=0.5\textwidth]{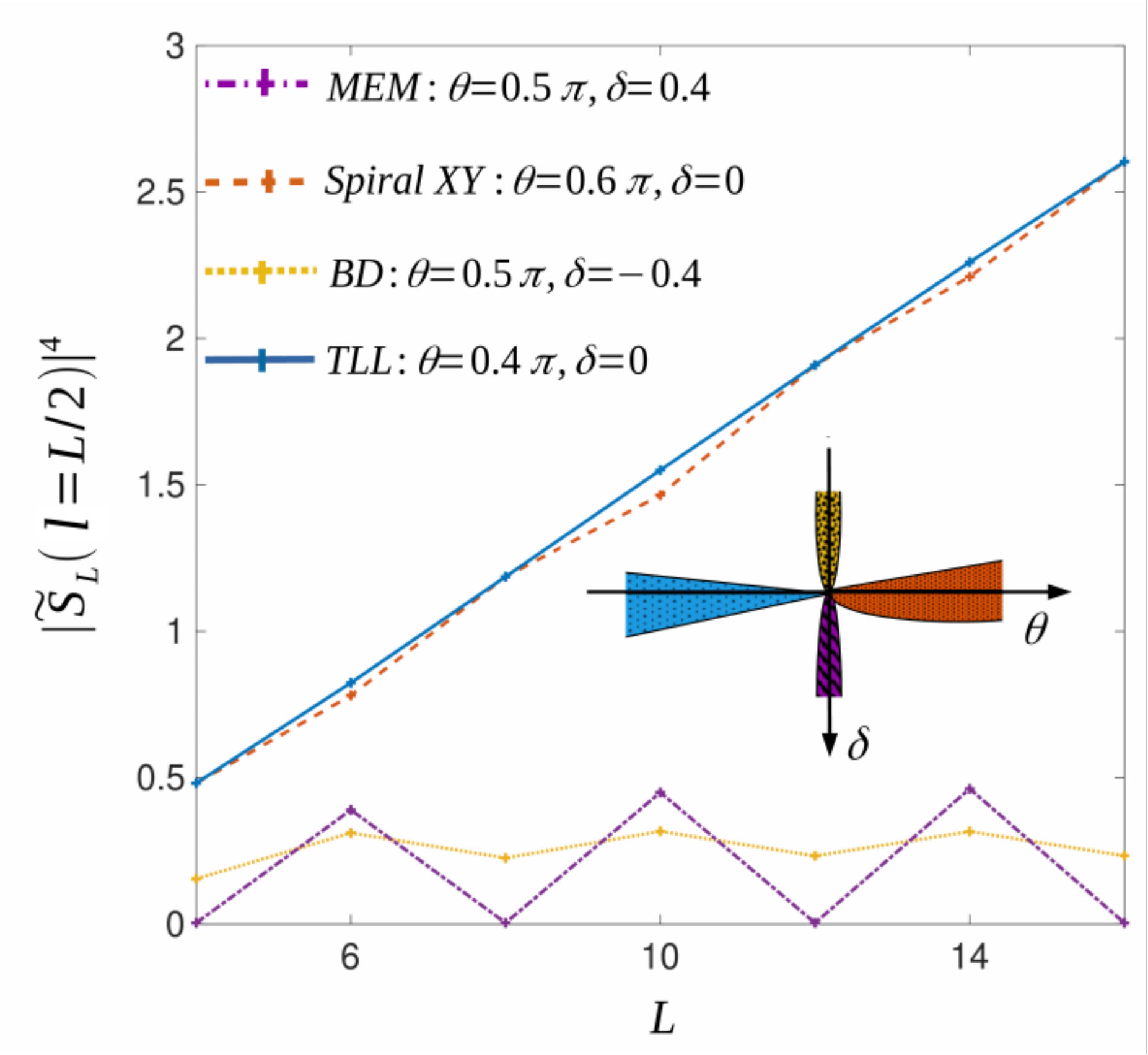}
  \caption{ Finite-size scaling of the maximum von Neumann entropy $\tilde{S}_L(l)$ of a subsystem with size $l=L/2$ of the Hamiltonian $\hat{H}_{KH}+V^\text{Ising}_\delta$. This entanglement measure is calculated for a range of system sizes $L$ in each of the relevant phases around the Kitaev point.   In the TLL and Spiral XY phases the von Neumann entropy scales with the size $\tilde{S}_L(L/2) \propto L^{1/4}$, whereas it oscillates with a periodicity of  $4$ in the other phases. We also note that (not shown here) the interactions in Eq.~\ref{majoranaA}, controlled by $\delta$, introduce large fluctuations in the finite-size scaling for both the TLL phase and the Spiral XY phase, thereby eventually breaking the TLL phase.
 }\label{Appendix_Fig2} 
 \end{figure*}

\begin{figure*}[h]
  \includegraphics[width=0.75\textwidth]{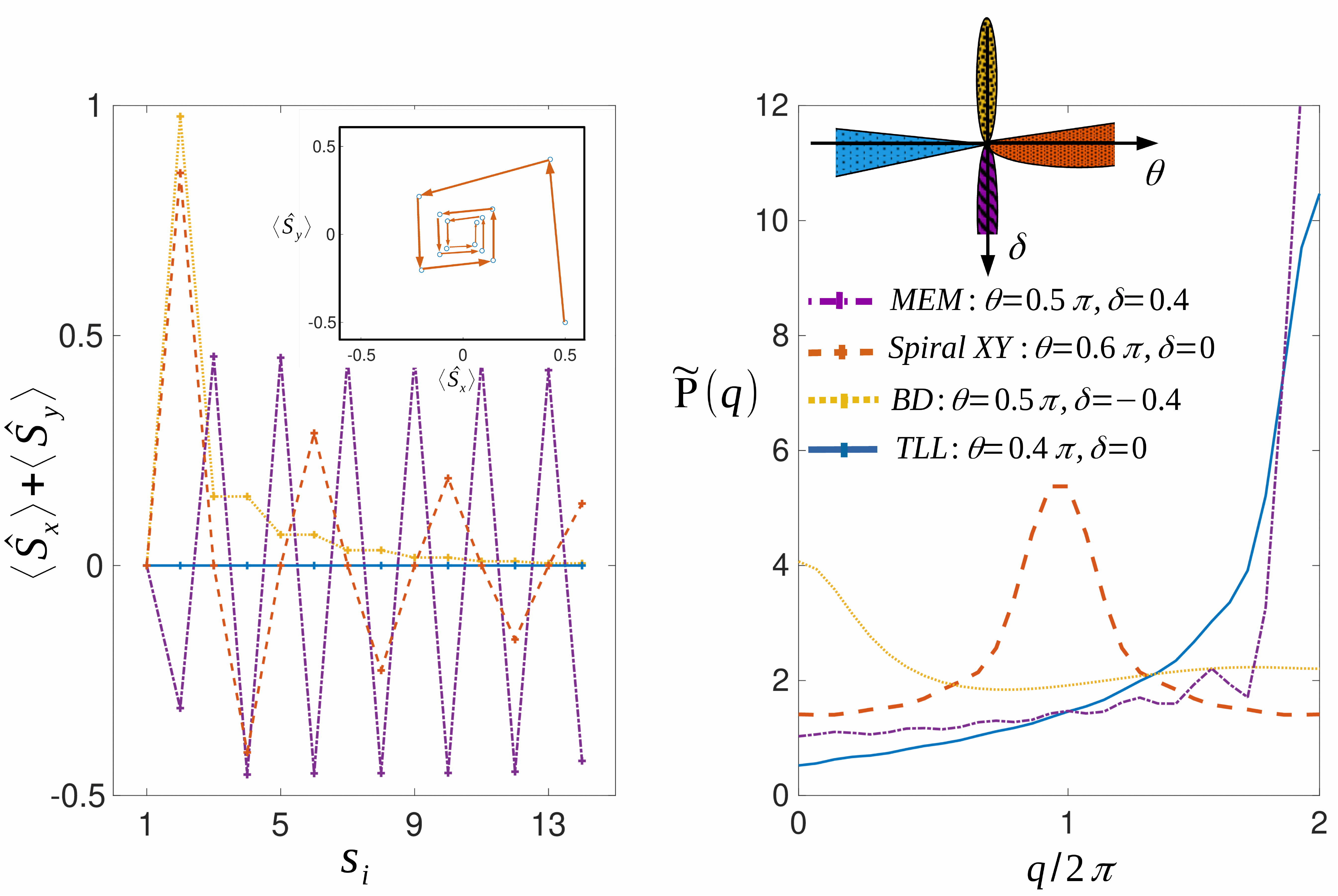}
  \caption{Left panel: Spin expectation value $\langle \hat{S}_x \rangle +\langle \hat{S}_y \rangle$ in four different phases of the Hamiltonian $\hat{H}_{KH}+V^\text{Ising}_\delta$. In the spiral-XY phase the unit cell still encompasses two sites, while the order parameter has periodicity of four lattice sites. The Majorana edge-mode phase has an antiferromagnetic structure in $S_x$ whereas no particluar structure is detected in the other phases. Inset: Spiral structure of the spiral-XY phase with arrows indicating the direction starting at the first site $i=1$.
  Right panel: Spin structure factors $\tilde{P}(q)=P^x(q)+P^y(q)$. The periodicity of the spiral-XY phase is indicated by a large peak at $q=\pi/2$.}\label{Appendix_Fig3} 
 \end{figure*}

\begin{itemize}
    \item  MEM phase ($\theta = \pi/2, \delta > 0 $): the x-components of the spin oscillate antiferromagnetically (Fig.~\ref{Appendix_Fig3}). The von Neumann entanglement entropy also oscillates but does not increase with system size (Fig.~\ref{Appendix_Fig2}).
    \item Spiral XY phase ($\theta >\pi/2, \delta \approx 0$): the spin expectation values indicate a spiral structure within the XY plane, with a spatial periodicity of four sites  (Fig.~\ref{Appendix_Fig3}). The entanglement entropy increases $\sim L^{1/4}$ and oscillates with periodicity four (Fig.~\ref{Appendix_Fig2}). 
    \item Gapped phase ($\theta= \pi/2, \delta < 0 $): the spin structure is ferromagnetic (Fig.~\ref{Appendix_Fig3}).  The entanglement entropy oscillates spatially with periodicity four, but does not increase with system size (Fig.~\ref{Appendix_Fig2}).
    \item TLL phase ($\theta < \pi/2, \delta \approx 0 $): the spin expectation value is zero throughout the chain (Fig.~\ref{Appendix_Fig3}).
    The entanglement entropy does not oscillate and scales according to $\sim L^{1/4}$ with total system size (Fig.~\ref{Appendix_Fig2}).
\end{itemize}

\section{Zero modes\label{asec_zeromodes}}
In this section we perform a derivation for the zero modes, i.e. low-energy modes which approximately commute with our Hamiltonian (see section~\ref{Zeromodes} in the main text). The presence of such modes largely explains the intriguing features of the autocorrelation simulations for the edge spins. 
Derivations follow the procedures in Refs.~\cite{kitaev2001unpaired,fendley2016strong,kemp2017long}, where higher-order corrections to a first-order approximation of the mode are sequentially added by commuting the mode with Hamiltonian. 
\subsection{Zero modes for $\hat{S}^y_2$}
We start with by considering the commuation between the operator $\hat{S}^y_2$, our first-order approximation for the zero mode, and the Hamiltonian. This gives
\begin{equation*}
\left[\hat{H}_{\text{MEM}},\hat{S}^y_2  \right]=i \frac{K}{K+\delta} \hat{S}^x_{\text{1}}\hat{S}^z_{\text{2}}
\end{equation*}
Since we want our mode to commute with the Hamiltonian, we offset this by introducing higher-order terms to the mode $\Psi_B^{(1)}=\mathcal{M}_B\hat{S}^z_{\text{1}}\hat{S}^y_{\text{2}}$,
$\Psi_B^{(2)}=4\hat{S}^x_{\text{1}}\hat{S}^x_{\text{2}}\hat{S}^y_{\text{3}}{K}/{(K+\delta)}$, 
so that
\begin{equation*}
\begin{split}
\left[\hat{H}_0, \Psi^{(1)}_B+\Psi^{(2)}_B \right]&=\left[\hat{S}^y_{\text{2}} \hat{S}^y_{\text{3}} , \Psi^{(2)}_B \right] \\ &=-i\frac{K}{K+\delta}\hat{S}^z_{\text{1}}\hat{S}^x_{\text{2}}
\end{split}
\end{equation*}
Now $\hat{V}$ and $\Psi^{(2)}_B$ do not commute, and produces a higher order term
\begin{equation*}
\begin{split}
\left[\hat{V},\Psi^{(2)}_B \right]&=\frac{K}{K+\delta}\left[\hat{S}^x_{\text{3}} \hat{S}^x_{\text{4}} , \Psi^{(2)}_B \right] \\
&=\left[\hat{S}^x_{\text{3}} \hat{S}^x_{\text{4}} , \hat{S}^x_{\text{1}}\hat{S}^x_{\text{2}}\hat{S}^y_{\text{3}} \right]4\left(\frac{K}{K+\delta}\right)^2 \\&=i\hat{S}^x_{\text{1}}\hat{S}^x_{\text{2}}\hat{S}^z_{\text{3}}\hat{S}^x_{\text{4}} \cdot 4\left(\frac{K}{K+\delta}\right)^2
\end{split}
\end{equation*}
which we offset by $\Psi_B^{(4)}=\hat{S}^x_{\text{1}}\hat{S}^x_{\text{2}}\hat{S}^z_{\text{3}}\hat{S}^z_{\text{4}} \hat{S}^y_{\text{5}} \cdot 4^2 (K/(K+\delta))^2$.
We continue by transforming to Pauli operators and finally get
\begin{equation}
\begin{split}
\Psi_B=&\sum^{L/2}_{\text{j=1}} \Psi_B^{(2j)}+\sum^{L/2}_{\text{j=1}}\Psi_B^{(2j-1)}\\
&=\mathcal{N}_{\text{e}}\sigma^y_2
+\mathcal{N}_{\text{e}}\sum^{L/2}_{\text{j=2}} \left(\frac{K}{K+\delta}\right)^{j-1} \sigma^y_{\text{2j-1}} \sigma^x_1 \sigma^x_2 \prod^{\text{2j-2}}_{\text{k=3}}\sigma^z_{\text{k}}\\ 
&+\mathcal{N}_{\text{o}}\sum^{L/2}_{\text{j=1}}  \mathcal{M}^{j-1}\sigma^y_{\text{2j}} \prod^{\text{2j-1}}_{\text{k=1}}\sigma^x_{\text{k}}\label{Zmodeap}
\end{split}
\end{equation}
Like in the case of for $\Psi_A$, all terms in $\Psi$ anticommute with each other, as well as with $\mathcal{G}^z$, so the normalization constant are the same as for both $\Psi_A$ and $\Psi_B$.

\subsection{Zero modes for $\hat{S}^z_1$}
As before, we note that $\Psi_C=\hat{S}^z_1$ commutes with $\hat{H}_0$ but not with $\hat{V}$, giving the commutation
\begin{equation*}
\left[\hat{H}_{\text{MEM}},\hat{S}^z_1  \right]=-i\frac{K}{K+\delta}\hat{S}^y_{\text{1}}\hat{S}^x_{\text{2}}
\end{equation*}
which we offset by introducing $\Psi_C^{(1)}=\mathcal{M}_C\hat{S}^z_{\text{1}}\hat{S}^y_{\text{2}}$,
$\Psi_C^{(2)}=4\hat{S}^y_{\text{1}}\hat{S}^z_{\text{2}}\hat{S}^y_{\text{3}}K/\left( K+\delta \right)$,
so that
\begin{equation*}
\begin{split}
\left[\hat{H}_0, \Psi^{(1)}_C+\Psi^{(2)}_C \right]&=4\frac{K}{K+\delta}\left[\hat{S}^y_{\text{2}} \hat{S}^y_{\text{3}} , \hat{S}^y_{\text{1}}\hat{S}^z_{\text{2}}\hat{S}^y_{\text{3}} \right] \\ &=i\frac{K}{K+\delta}\hat{S}^y_{\text{1}}\hat{S}^x_{\text{2}}
\end{split}
\end{equation*}
giving for the next order
\begin{equation*}
\begin{split}
\left[\hat{V},\Psi^{(2)}_C \right]&=\frac{K}{K+\delta}\left[\hat{S}^x_{\text{3}} \hat{S}^x_{\text{4}} , \Psi^{(2)}_B \right] \\
&=4\left[\hat{S}^x_{\text{3}} \hat{S}^x_{\text{4}} , \hat{S}^y_{\text{1}}\hat{S}^z_{\text{2}}\hat{S}^y_{\text{3}} \right]\left(\frac{K}{K+\delta}\right)^2 \\&=i\hat{S}^y_{\text{1}}\hat{S}^z_{\text{2}}\hat{S}^z_{\text{3}}\hat{S}^x_{\text{4}} \cdot 4\left(\frac{K}{K+\delta}\right)^2
\end{split}
\end{equation*}
which is offset by $\Psi_C^{(4)}=-\hat{S}^y_{\text{1}}\hat{S}^z_{\text{2}}\hat{S}^z_{\text{3}}\hat{S}^z_{\text{4}} \hat{S}^y_{\text{5}} \cdot 4^2 (K/(K+\delta))^2$.
We continue by transforming to Pauli operators giving
\begin{equation}
\begin{split}
\Psi_C=&\sum^{L/2}_{\text{j=1}} \Psi_C^{(2j)}+\sum^{L/2}_{\text{j=1}}\Psi_C^{(2j-1)}\\
&=\mathcal{N}_{\text{e}}\sigma^z_1
+\mathcal{N}_{\text{e}}\sum^{L/2}_{\text{j=2}} (-1)^j \left(\frac{K}{K+\delta}\right)^{j-1} \sigma^y_{\text{2j-1}} \sigma^y_1  \prod^{\text{2j-2}}_{\text{k=2}}\sigma^z_{\text{k}}\\ 
&+\mathcal{N}_{\text{o}}\sum^{L/2}_{\text{j=1}}  \mathcal{M}^{j-1}\sigma^y_{\text{2j}} \prod^{\text{2j-1}}_{\text{k=1}}\sigma^x_{\text{k}}
\end{split}
\end{equation}
Again, all terms in $\Psi$ anticommute with each other so the normalization constant are the same as for both $\Psi_A$ and $\Psi_B$. Also, each term has an odd number of $\sigma^z$ so $\Psi_C$ anticommutes with $\mathcal{G}^x$ and $\mathcal{G}^y$.

\section{Spin coherence\label{asec_zeromodedynamics}}
Here, we derive the influence of zero modes on the dynamical development of single spins in our model. A schematic derivation of the autocorrelation is performed in the main text in section~\ref{zeromodedynamics}, and this section is included as a complement.  
We start by noting that
\begin{equation}
\begin{split}
    \left[ H, \Psi_A \right] \approx 0 \\
    \{ \mathcal{G}^z, \Psi_A \} = 0\\
    \mathcal{G}^z \mathcal{G}^z=\Psi_A^2=1
\end{split}
\end{equation}
and proceed to write down the autocorrelation function $ \Gamma^y_{1}(t)$ for an eigenstate $|S^y\rangle$, with corresponding eigenvalue $s^y_1$, of the edge spin Pauli matrix $\sigma^y_1$ along direction $y$
\begin{equation}
\begin{split}
    \Gamma^y_{1}(t)=& \langle S^y |\hat{S}^{y}_{1}(t) \hat{S}^{y}_{1}(t=0)| S^y \rangle\\=(1/4)&\langle S^y |e^{-i\hat{H}t} \sigma^y_1 e^{-i\hat{H}t}\sigma^y_1 | S^y \rangle \\=&\frac{s^y_1}{4} \sum_{n,m} \langle S^y|n\rangle \langle n| e^{-i\hat{H}t} \sigma^y_1 e^{-i\hat{H}t} |m\rangle \langle m| S^y \rangle 
\\=&\frac{s^y_1}{4}\sum_{n,m} e^{-it\left( E_m-E_n\right)} \langle S^y|n\rangle \langle n| \sigma^y_1|m\rangle \langle m| S^y \rangle
\end{split}
\end{equation},
where $  |n\rangle, |m\langle$ are eigenstates of the Hamiltonian.
Since $\Psi_A$ almost commutes with the Hamiltonian $H$ we may divide all energy states into two sectors denoted by positive or negative signs, corresponding positive or negative eigenvalues of $\Psi_A$ so that
\begin{equation}
\begin{split}
    &H|n^A_{\pm}\rangle \approx E_n|n^A_{\pm}\rangle\\
    &\Psi_A|n^A_{\pm}\rangle= \pm|n^A_{\pm}\rangle
\end{split}
\end{equation}
We can now re-write the autocorrelation function for the new indices
\begin{align*}
    \Gamma^y_{1}(t)=& \frac{s^y_1}{4} \sum_{n^A,m^A} \left( \langle S^y \left(|n^A_{+}\rangle \langle n^A_{+}|+|n^A_{-}\rangle \langle n^A_{-}|\right) e^{-i\hat{H}t}  \right.
   \\& \left. \sigma^y_1 e^{-i\hat{H}t} \left(|m^A_{+}\rangle \langle m^A_{+}|+|m^A_{-}\rangle \langle m^A_{-}|\right) S^y \rangle \right)\\
\end{align*}
For long times $t$ and large system size $L$, terms with $n^A\neq m^A$ add up incoherently while terms with $n^A=m^A$ add up coherently, so the double sum may be approximated
\begin{equation}
\begin{split}
    \Gamma^y_{1}(t) \approx \frac{s^y_1}{4} \sum_{n^A} \cdot \left(T_A+T_B+T_C+T_D\right)
\end{split}
\end{equation}
with
\begin{equation*}
\begin{split}
    T_1=\langle S^y |n^A_{-} \rangle \langle n^A_{-}| \sigma^D_1 |n^A_{-} \rangle \langle n^A_{-}|S^y\rangle  \\
        T_2=\langle S^y |n^A_{-} \rangle \langle n^A_{-}| \sigma^y_1 |n^A_{+} \rangle \langle n^A_{+}|S^y\rangle  \\
            T_3=\langle S^y |n^A_{+} \rangle \langle n^A_{+}| \sigma^y_1 |n^A_{-} \rangle \langle n^A_{-}|S^y\rangle  \\
                T_A=\langle S^y |n^A_{+} \rangle \langle n^A_{+}| \sigma^y_1 |n^A_{+} \rangle \langle n^A_{+}|S^y\rangle  \\
\end{split}
\end{equation*}
For $T_2$ and $T_3$ we exploit the spin-flip operator in Eq.~\ref{spinflip} which anticommutes with $\Psi_A$ so that $\mathcal{G}^z |n^A_{\pm} \rangle =  |n^A_{\mp} \rangle$ and we get
\begin{equation}
\begin{split}
    T_2=&\langle S^y |n^A_{-} \rangle \langle n^A_{+}|S^y\rangle \cdot  \langle n^A_{-}| \sigma_1^y |n^A_{+} \rangle\\
    =&\langle S^y |n^A_{-} \rangle \langle n^A_{+}|S^y\rangle \cdot \frac{\left( \langle n^A_{-}|\sigma_1^y|n^A_{+} \rangle + \langle n^A_{+}|\sigma_1^y|n^A_{-} \rangle\right)}{2} \\
    =&\langle S^y |n^A_- \rangle \langle n^A_+|S^y\rangle \cdot  \langle n^A_+|\mathcal{G}^z\sigma_1^y +\sigma_1^y\mathcal{G}^z|n^A_+ \rangle /2\\
    =&\langle S^y |n^A_- \rangle \langle n^A_+|S^y\rangle \cdot  \langle n^A_+|\{ \sigma_1^y,\mathcal{G}^z\}|n^A_+ \rangle /2\\
    T_3=&\langle S^y |n^A_+ \rangle \langle n^A_-|S^y\rangle \cdot  \langle n^A_+| \{ \sigma_1^y, \mathcal{G}^z  \} |n^A_+ \rangle/2\\
\end{split}
\end{equation}
For $T_1$ and similarly for $T_4$ we get
\begin{equation}
\begin{split}
    &T_1/|\langle S^y |n^A_- \rangle |^2=  
    \langle n^A_-| \sigma^y_1 |n^A_- \rangle\\
    &=\langle n^A_+|\mathcal{G}^z \sigma^y_1 \mathcal{G}^z|n^A_+ \rangle\\
    &= \left(  \langle n^A_+|\mathcal{G}^z \sigma^y_1 \mathcal{G}^z|n^A_+ \rangle +
    \langle n^A_+|\mathcal{G}^z \sigma^y_1 \mathcal{G}^z|n^A_+ \rangle \right)/2\\
    &= \left(  
    \langle n^A_+|\Psi_A \mathcal{G}^z \sigma^y_1 \mathcal{G}^z|n^A_+ \rangle +
    \langle n^A_+|\mathcal{G}^z \sigma^y_1 \mathcal{G}^z\Psi_A |n^A_+ \rangle \right)/2\\
    &=  
    \langle n^A_+|\{ \Psi_A, \mathcal{G}^z \sigma^y_1 \mathcal{G}^z \} |n^A_+ \rangle/2\\
    &T_4/|\langle S^y |n^A_+ \rangle |^2=  \langle n^A_+| \{ \Psi_A, \sigma^y_1  \} |n^A_+ \rangle/2
\end{split}
\end{equation}

\section{Spin beating\label{asec_spinbeating}}
Here, we derive the origin for the beating mechanism in the temporal evolution of edge spins, as discussed in section~\ref{sec_beating} of the main text. See section~\ref{sec_beating} for definitions and background.
We use the same technique as in the previous section, but with an additional trick to derive the short-time dynamics in the autocorrelation function. 
We start with
 \begin{equation} \hat{H}=\hat{H'} + \hat{H}^{\text{P}}
 \end{equation}
 where
 \begin{equation}
 \begin{split}
 &\hat{H'}=\hat{H}_{KH}(\theta=\pi/2)\\
 &\hat{H}^{\text{P}}= \sum_{\text{i=1}}^{L} \mathcal{A}_i \hat{S}^y_{\text{i}}S^y_{\text{i+1}}
 \end{split}
 \end{equation}
Here we could reduce the general perturbation to a pure Ising term with strength $\delta$ by putting
  \begin{equation*}\hat{V}^{\text{Ising}}_{\delta} = \hat{H}^{\text{P}} \quad\forall \mathcal{A}_i=\delta
 \end{equation*}
 but now we keep the more general form of the perturbation.
 We note that two terms in the Hamiltonian commute
  \begin{equation*}  \left[ \hat{H'},\hat{H}^{\text{P}} \right]=0 
 \end{equation*} and we proceed to study the time evolution 
\begin{equation}
\begin{split}
    \Gamma^z_{1}(t)=& \frac{s^z_1}{4} \sum_{n,m} \langle S^z|n\rangle \langle n| e^{-i(\hat{H'}+\hat{H}^{\text{P}})t} \sigma^z_1 e^{i(\hat{H'}+\hat{H}^{\text{P}})t} |m\rangle \langle m| S^z \rangle 
\\=&\frac{s^z_1}{4} \sum_{n,m} e^{it\left( E_m-E_n\right)} \langle S^z|n\rangle \langle n| \sigma^z_1(t)|m\rangle \langle m| S^z \rangle
\end{split}
\end{equation}
Here, we can again take the long time limit and divide into positive and negative eigenvalues of $\Psi_C$, but we first turn our attention to the time dependent matrix element defined as
\begin{equation}
  \sigma^z_1(t) = e^{-i\hat{H}^{\text{P}}t} \sigma^z_1 e^{i\hat{H}^{\text{P}}t}
\end{equation}
We note that all terms in $\hat{H}^{\text{P}}$ except the first one commute with $\sigma^z_1$, so that
\begin{equation}
  \sigma^z_1(t)=e^{-it\mathcal{A}_1\sigma^y_1\sigma^y_1}\sigma^z_1
 e^{it\mathcal{A}_1\sigma^y_1\sigma^y_1}
 \label{HP}
\end{equation}
We now expand the exponentials so that 
\begin{equation}
\begin{split}
     &e^{it\mathcal{A}_1\sigma^y_1\sigma^y_2}= 1 +(it\mathcal{A}_1)(\sigma^y_1\sigma^y_2)+ \frac{ (it\mathcal{A}_1)^2(\sigma^y_1\sigma^y_2)^2}{2!}\\
     &\frac{(it\mathcal{A}_1)^3(\sigma^y_1\sigma^y_2)^3}{3!}+\frac{(it\mathcal{A}_1)^4(\sigma^y_1\sigma^y_2)^4}{4!}+...\\
     &=1-\frac{(t\mathcal{A}_1)^2}{2!}+\frac{(t\mathcal{A}_1)^4}{4!}-...\\
     &+i\sigma^y_1\sigma^y_2\left(t\mathcal{A}_1 -\frac{(t\mathcal{A}_1)^3}{3!}+... \right)\\
     &=\cos{\mathcal{A}_1t} +i \sigma^y_1\sigma^y_2 \sin{\mathcal{A}_1t}
\end{split}
\end{equation}
We use this to expand Eq.~\ref{HP} yielding
\begin{equation}
\begin{split}
  \sigma^z_1(t)&=\sigma^z_1 \cos^2{\mathcal{A}_1t}\\
 &+ 2 \cos{\mathcal{A}_1t}\sin{\mathcal{A}_1t} \sigma^x_1\sigma^y_2\\
 &-\sigma^z_1\sin^2{\mathcal{A}_1t}
\end{split}
\end{equation}
Let us now compare to the derivation in the previous section and employ the substitutions $ \sigma^y_1\rightarrow \sigma^z_1(t)$, $\mathcal{G}^z\rightarrow\mathcal{G}^x$, $\Psi_A\rightarrow\Psi_C$, $\langle S^y|\rightarrow\langle S^z|$ and $\langle n^A_\pm| \rightarrow \langle n^C_\pm|$. The autocorrelator can now be expressed as
\begin{equation}
\begin{split}
    \Gamma^z_{1}(t) \approx \frac{s^z_1}{4} \sum_{n^C} \cdot \left(T_A+T_B+T_C+T_D\right)
\end{split}
\end{equation}
where 
\begin{equation}
\begin{split}
    T_2=&\langle S^z |n^C_- \rangle \langle n^C_+|S^z\rangle \cdot  \langle n^C_+|\{ \sigma_1^z(t),\mathcal{G}^x\}|n^C_+ \rangle /2\\
    T_3=&\langle S^z |n^C_+ \rangle \langle n^C_-|S^z\rangle \cdot  \langle n^C_+| \{ \sigma_1^z(t), \mathcal{G}^x  \} |n^C_+ \rangle/2\\
\end{split}
\end{equation}
All terms of $\sigma^z_1(t)$ anticommute with $\mathcal{G}^x$, so that $T^2=T^3=0$ and
\begin{equation}
\begin{split}
    &T_1=-|\langle S^z |n^C_- \rangle |^2
    \langle n^C_+|\{ \Psi_C, \sigma^z_1(t)\} |n^C_+ \rangle/2\\
    &T_4=|\langle S^z |n^C_+ \rangle |^2 \langle n^C_+| \{ \Psi_C, \sigma^z_1(t)  \} |n^C_+ \rangle/2
\end{split}
\end{equation}
The second term in $\sigma^z_1(t)$ anticommutes with $\Psi_C$, so that we are left with
\begin{equation}
\begin{split}
    &T_1+T_4=\left(|\langle S^z |n^C_+\rangle |^2-|\langle S^z |n^C_- \rangle |^2\right)
   \frac{ \langle n^C_+|\{ \Psi_C, \sigma^z_1(t)\} |n^C_+ \rangle}{2}\\
    &= (\cos^2{\mathcal{A}_1t} - \sin^2{\mathcal{A}_1t})\left(|\langle S^z |n^C_+\rangle |^2-|\langle S^z |n^C_- \rangle |^2\right)\cdot M
\end{split}
\end{equation}
where 
\begin{equation}
    M= \langle n^C_+|\{ \Psi_C, \sigma^z_1\} |n^C_+ \rangle\approx \mathcal{N}_e + C 
\end{equation}
By setting $\mathcal{A}_1=\delta$ the result explains the beating mechanism in section~\ref{sec_beating} of the main text.

\section{Quantum simulation with trapped ions\label{asect-quantumsim}}
This section serves as a complement to the setup of the quantum simulation in section~\ref{sect-quantumsimulation}.
For both the XX and ZZ interaction implementation schemes in our setup we aim to virtually excite \textit{only} transverse phonons in the strongly confined direction $y$. This means that the angular frequencies of the lasers must lie far from resonance with the zig-zag planar phonons. This is ensured by a hierarchy of trapping frequencies, where $\omega_x$ differs from $\omega_y$ by a large factor. We specifically consider a Paul trap with axial centre-of-mass frequency $\omega_z\sim 80$ kHz and transverse frequencies $\omega_x\sim 1.5$ MHz, $\omega_y \sim 10$ MHz. The trappings in $z$ and $x$ are similar to the parameters employed in Ref.~\cite{zhang2017observation}, whereas the trapping in $y$ is roughly two times larger.
We assume $N\sim 70$ $^{171}\text{Yb}^+$ ions which gives a typical axial distance of $\sim 1.5 \smallskip$ $\mu$m at the centre of the trap~\cite{james2000quantum}. 
Different types of ions, including singly excited alkaline-earth ions, could be used as long as they allow for both the XX and ZZ interactions considered here. The ions must also be able to be efficiently trapped with typical ion-distances $l_z \sim 1-10 $ $\mu$m~\cite{bermudez2012quantum}.

For the $\Lambda$ scheme we assume bichromatic laser with angular frequencies $\omega^\text{a}_1= \omega^\text{a}-\omega^\text{a}_l$ and $\omega^\text{a}_2= \omega^\text{a} + \omega^\text{a}_l$, with effective Raman wave vector $\vec{k}^\text{a}_l = \vec{k}^\text{a}_1 - \vec{k}^\text{a}_2$, corresponding to an effective wavelength   $\lambda_l=2\pi/|\vec{k}_l| \sim 400$ nm. 
 The optical beatnote frequency between the Raman lasers is roughly matched to the transverse frequency so that $\omega^\text{a}_l \approx \omega_y$, with sideband detunings $\omega^\text{a}_\Delta= \omega^\text{a}_l-\omega_y = 480$ kHz. 
Assuming $\lambda^\text{a}_l=400$ nm, 
the nearest-neighbour ion distance typically corresponds to a handful of wavelengths, which is reasonable for our purposes~\cite{bermudez2012quantum}.
The full form of the effective ZZ interaction is given by
\begin{equation}
 \begin{split}
     {V}^\text{a}_\text{ij}= -\Omega_\text{a}^2 &\frac{\hbar (\sin{\varphi_a})^2 |k^\text{a}_{l}|^2}{8m}\\
     &\cdot\sum_\text{p} \frac{M^\text{y}_\text{i,p} M^\text{y}_\text{j,p} \cos{\left( \vec{k}^\text{a}_{l} \cdot \tilde{\vec{r}}^0_\text{ij} \right)} }{ \omega^\text{y}_\text{p} \left(\omega^\text{y}_\text{p}- \omega^\text{a}_{l}\right)} 
 \end{split}
 \label{ZZ_int}
 \end{equation}
 with vibrational mode eigenvectors $M^\text{y}_\text{p}$ and mode frequencies $\omega^\text{y}_\text{p}$ along the strongly confined y-direction~\cite{bermudez2012quantum}. 
The effective two-photon Rabi frequency is given by $\Omega_\text{a}=\left(\Omega_{1,\downarrow}\Omega_{2,\downarrow}^*+\Omega_{1,\uparrow}\Omega_{2,\uparrow}^* \right)/2\mathcal{D}$.

In our simulations, we approximate the distances between the rungs and legs in the active region of the trap to be homogeneous. The actual variations in distances introduce additional errors, but can be remedied by tailoring the shape of the trapping potential~\cite{skrevet2018effect}, and in principle also by adding more inactive ions to the setup. The angle in our showcase simulation is set at $\varphi_a=0.5164$ rad, which roughly eliminates the inter-rung interaction for our choice of parameters for the trap and fields.

For the XX interaction we consider transitions between the internal levels and we again use bichromatic driving with angular frequencies $\omega^\text{b}_1= \omega^\text{b}-\omega^\text{b}_l$ and $\omega^\text{b}_2= \omega^\text{b} + \omega^\text{b}_l$, with an effective Raman wave vector $\vec{k}^\text{b}_l$, corresponding to wavelengths  $\lambda^\text{b}_l=2\pi/|\vec{k}^\text{b}_l| \sim 400$ nm. 
 The frequency of the lasers is roughly matched to the transverse frequency so that $\omega^\text{b}_l \approx \omega_y$ with sideband detunings $\omega^\text{b}_\Delta= \omega^\text{b}_l-\omega_y = 540$ kHz.

\subsection{Residual interactions~\label{residuals}}
As noted in section~\ref{sect-quantumsimulation} of the main text, the relative size of the residual interactions, which induce effective long-range interactions, is a big hindrance in simulating the Majorana beating mechanism. 
For $K=1$ and $\delta_\text{eff}\sim 1$, the spectral gaps associated with the residual interactions are larger than the Majorana finite-size gap $\Delta_L$ already at very small system sizes $L\sim6$. Therefore, the observable spin oscillations of the edge sites decay due to the residual interactions before any beating can be observed.
It is however possible to reduce the relative size of the residual interactions by additional optical fields. 
For example, one may supplement the lasers which generate the XX interaction 
by additional laser fields at different sideband detunings.  
The additional fields should have smaller detunings relative to the first fields so that they effectively realize a longer-range interaction. 
By correctly matching the effective Rabi frequencies for the two interaction terms (now with different ranges) one may effectively cancel most of the next-nearest neighbour residual interactions. 
These additional fields for the XX interactions can reduce the most "destructive" residual terms by an order of magnitude. 

For the  ZZ interactions, which depend on the effective wavevector of the fields, a pair of fields with effective wavevector in the $zy$-plane (see Fig.~\ref{Fig5}) can be exploited.
By adding a second field at smaller detuning compared to the first, we can reduce the residuals by a factor $\sim 4$. This can be done by carefully tuning the angles, controlling the effective interactions via the expression~\ref{ZZ_int}, and modifying the Rabi frequency to cancel the largest residuals.

\end{document}